\documentclass[11pt]{article}
  \usepackage{epsf}
\newcommand{\mpr}{\,\scriptscriptstyle\parallel}
\newcommand{\mpp}{\,\scriptscriptstyle\perp}

\newcommand{\ppp}{p_{\mpp}}
\newcommand{\ppr}{p_{\mpr}}
\newcommand{\popp}{p_{\mpp\,0}}
\newcommand{\popr}{p_{\mpr\,0}}

\begin{document}
  \title{ \vspace{-12mm}
         \bf Magnetic reconnection and topological trigger
         in physics of large solar flares}

  \author{Boris V. Somov   \vspace{1.5mm} \\
          {\em Astronomical Institute and Faculty of Physics,
          Moscow State University,}\\
          {\em Moscow 119992, Russian Federation}\\
          {\em somov@sai.msu.ru} \\}

  \date{}
  \maketitle

\noindent
{\bf Abstract}
\begin{center}
\parbox{150mm}{ %%%%
Solar flares are accessible to a broad variety of observational
methods to see and investigate the {\em magnetic reconnection\/}
phenomenon in high-temperature strongly-magnetized plasma of the
solar corona.
An analysis of the topological peculiarities of magnetic field in
active regions shows that the {\em topological trigger\/} effect
is necessary to allow for in order to construct models for large
eruptive flares.

\noindent
\hspace*{6mm}
The topological trigger is not a resistive instability which
leads to a change of the topology of the field
configuration from pre- to post reconnection state.
On the contrary, the topological trigger is a quick change of the
global topology, which dictates the fast reconnection of col\-lisional
or collision\-less nature.

\noindent
\hspace*{6mm}
The current state of the art and development potential of the theory
of collision\-less reconnection in the strong magnetic fields related
to large flares are briefly reviewed.
Particle acceleration is considered in collapsing magnetic traps
created by reconnection.
In order to explain the formation of coronal X-ray sources,
the Fermi acceleration and betatron mechanism are simultaneously
taken into account analytically in a collision\-less approximation.
Finally, the emphasis is on urgent unsolved problems of solar flare
physics.
} %%%%
\end{center}

%
%%% Section 1 %%%%%%%%%%%%%%   %%%%%%%%%%%%%%%%%%%%%%%%%%%%%%%%%%%%%%%%%
%
\section{The role of magnetic fields in flares}
   \label{sec:Romf}

%
%%%%%%%%%%%%%%%% Sub-section 1.1 %%%%%%%%%%%%%%%%%%%%%%%%%%%%%%%%%%%%%%%
%
\subsection{Basic questions}
   \label{sub:051211a}

Frequent observations of solar flares became available in the 1920s.
Early studies showed that flares were associated with magnetic fields.
Estimates of the energy required to power large flares led to the
conclusion that flares must be electromagnetic in origin.
Step by step it became more clear that a flare is the result
of reconnection of magnetic fields.
However there were some objections to the hypothesis
that the energy of a flare could be stored in the form of a
magnetic field of {\em reconnecting current layers\/}
(RCLs).

{\bf (1)}
It was claimed that measurements of photospheric magnetic
fields do not demonstrate an unambiguous relation between flares and
the changes of the fields.
More exactly, the changes in question are those that occur
{\em immediately before\/} a flare to create it.
These changes were supposed to be the cause
of a flare.

{\bf (2)}
Next objection was related to the time of dissipation of the
magnetic field in a volume that would contain the energy necessary
for a flare.
If this time is estimated in a usual way as the diffusion time in a
solar plasma of a finite conductivity, then it is too long compared
with the observed duration of a flare.

{\bf (3)}
The third objection was the most crucial one: the observers had
never seen real RCLs in solar flares.

\vspace{0.3mm}

Starting from
Severny~[1],
%%% Severny~(1964), %%%
solar observers have been studying flare-related changes in
photospheric magnetic fields to provide information
how an active region (AR) stores and releases its energy (see
Lin et al.~[2],
Wang~[3]).
%%% Lin et al.,~1993; %%%
%%% Wang,~1999). %%%
However the role of photospheric fields is still an area of ongoing
research
(Section~\ref{sub:051211c}).
What are the answers of the reconnection theory to the objections
listed above?

%
%%%%%%%%%%%%%% Sub-section 1.2 %%%%%%%%%%%%%%%%%%%%%%%%%%%%%%%%%%%%%%%%%
%
\subsection{Concept of magnetic reconnection in solar flares}
   \label{sub:051211b}

According to contemporary views, the principal flare process is
contingent on the accumulation of the {\it free magnetic energy\/}
in the corona.
By `free' we mean the surplus energy above that of a potential
magnetic field
% (1)
\begin{equation}
    B_{\alpha} ( {\bf r} )
  = \, \frac{ \partial \psi }{ \partial r_{\alpha} } \, .
   \label{071130}
\end{equation}
Here $ \psi  ( {\bf r} ) $ is the potential of the field,
the index~$ \alpha = 1,2,3 $.
The potential field has the sources (sunspots, background
fields) in the photosphere.

The free magnetic energy is related to the electric currents in the
corona.
A solar flare corresponds to rapid changes of these currents.
So we distinguish between two processes: a slow accumulation of energy
and its fast release, a flare.
%
%%%%%% Figure 1 %%%%%%   %%%%%%%%%%%%%%%%%%%%%%%%%%%%%%%%%%%%%%%%%%%%%%%
%
\begin{figure} [h]
   \epsfxsize=105mm
   \centerline{\epsfbox{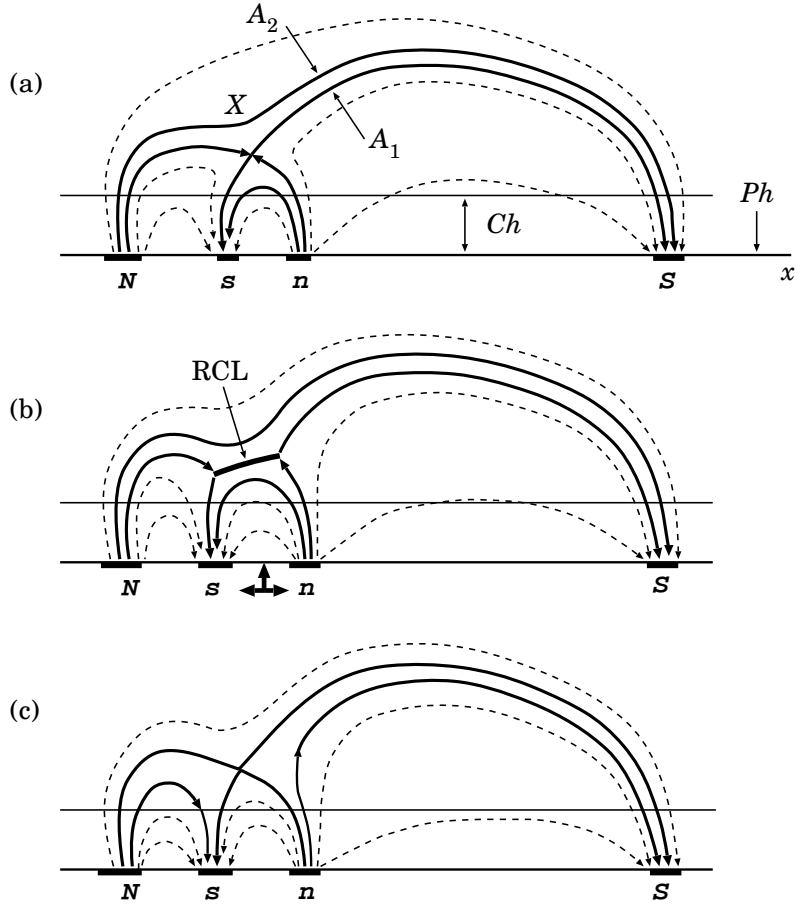}}
   \caption{The classical 2D cartoon of magnetic reconnection in a
      solar flare.
      Three states of the field:
      (a) the initial one,
      (b) the {\it pre-reconnection\/} state,
      (c) the final state after reconnection. }
   \label{fig1}
\end{figure}
Let us see them in a classical example,
the evolution of the quad\-ru\-pole configuration of sunspots
(Figure~\ref{fig1}).
%
%%% Figure 1 %%%
%
The sunspots of pairwise opposite polarity are shown: $ N $ and~$ S $
represent a bipolar group of sunspots in an AR,
$ n $ and~$ s $ model a new emerging flux.
All sunspots are placed along the axis~$ x $ in the photospheric
plane~$ Ph $ at the bottom of the chromosphere~$ Ch $.

The field line~$ A_1 $ in
Figure~\ref{fig1}a
%
%%% Figure 1 (a) %%%
%
is the se\-pa\-rat\-rix line of the initial
state~(a), this line will reconnect first.
$ X $ is the zeroth point of the field at the initial state, here
the RCL is created at the state~(b).
The field line~$ A_2 $ is the se\-pa\-rat\-rix of the final
state~(c) or the last reconnected field line.
Therefore $ \delta A =  A_2 -  A_1 $ is the reconnected magnetic
flux.

Three solid arrows under the photosphere~$ Ph $ in
Figure~\ref{fig1}b
%
%%% Figure 1 (b) %%%
%
show a slow emergence of a new magnetic flux (the sunspots~$ n $
and~$ s $).
The sunspots have been emerged but the field lines do not start to
reconnect.
More exactly, they reconnect {\em too slowly\/} because of
high conductivity of plasma in the corona.
In the first approximation, we neglect this reconnection.

In general, the redistribution of fluxes appears as a result of the
slow changes of the field sources in the photosphere.
These changes can be either the emergence of a new flux from
below the photosphere
(Figure~\ref{fig1})
%
%%% Figure 1 %%%
%
or other flows of photospheric plasma, in particular the
shear flows, inhomogeneous horizontal flows along the
neutral line of the photospheric magnetic field.
For this reason, an actual ({\em slow\/} and {\em fast\/})
reconnection in the corona is always a three-dimensional
(3D) phenomenon
(Section~\ref{sec:Tdri}).
%
%%% Section 2 %%%
%

%
%%% Section 1.3 %%%%%%%%%%%%%%%%%%%%%%%%%%%%%%%%%%%%%%%%%%%%%%%%%%%%%%%%
%
\subsection{Some new results of the magnetic field observations}
   \label{sub:051211c}

Let us come back to objection~{\bf (1)} in
Section~\ref{sub:051211a}
%
%%% Sub-section 1.1 %%%
%
to the reconnection theory of flares.
According to the theory, the free energy is related to the
current~$ J $ inside the RCL.
A flare corresponds to rapid changes of this current.
However the magnetic flux through the photosphere~$ Ph $
(Figure~\ref{fig1})
%
%%% Figure 1 %%%
%
changes only little over the whole area of a flare
during this process, except in some particular places, for example,
between close sunspots~$ N $ and~$ s $.

Sunspots in the photosphere
are weakly affected by a flare because the plasma
in the photosphere is almost $ 10^9 $ times denser than the plasma in
the corona.
It is difficult for disturbances in the tenuous
corona to affect the extremely massive plasma in the photosphere.
Only small perturbations penetrate into the photosphere.

The same is true in particular for the vertical component of the
field.
The photospheric magnetic field changes a little during a flare over
its whole area.
As a consequence, after a flare the large-scale structure in the
corona can remain free of noticeable changes, because it is determined
mainly by the potential part
of the field above the photosphere.
More exactly, even being disrupted, the structure will
come to the potential configuration corresponding to the
post-flare position of the photospheric sources.

On the other hand, in the ``Bastille-day'' flare on~2000 July~14 and
in some other large flares, it was possible to detect the real
changes in a sunspot structure just after a flare.
The outer pen\-umber fields became more vertical due to reconnection
in the corona during a flare
(Liu et al.~[4],
Wang et al.~[5]).
%%% (Liu et al.,~2005; %%%
%%% Wang et al.,~2005). %%%
One can easily imagine such changes by considering
Figure~\ref{fig1}
%
%%% Figure 1 %%%
%
between sunspots~$ N $ and~$ s $.

Sudol and Harvey~[6]
%%% Sudol and Harvey~(2005) %%%
used the magneto\-grams to characterize the chan\-ges in the vertical
component of photospheric field during 15 large flares.
An abrupt, significant, and persistent change occurred in
at least one location within the flaring AR during each event after
its start.
Among possible interpretations,
Su\-doh and Harvey
favor one in which the field changes result from the
pen\-umber field relaxing upward by reconnecting magnetic field above
the photo\-sphere.
This interpretation is similar to than one given by
Liu et al.~[4] and
Wang et al.~[5].
%%% Liu et al.~(2005) and %%%
%%% Wang et al.~(2005). %%%

As for objection~{\bf (2)} to the hypothesis of
accumulation of energy in the form of magnetic field of
current layers, the rapid dissipation of
the field necessary for a flare is explained by the theory
of super-hot turbulent-current layers
(Section~\ref{sec:BPoH}).
%
%%% Section 4 %%%
%
In what follows, we advocate that the potential part of magnetic
field in the solar corona determines a large-scale structure and
properties of active regions, while the reconnecting current layers
determine energetics and dynamics of flares.

%
%%% Section 2 %%%   %%%%%%%%%%%%%%%%%%%%%%%%%%%%%%%%%%%%%%%%%%%%%%%%%%%%
%
\section{Three-dimensional reconnection in flares} %%%%%%%%%%%%%%%%%%%%%
   \label{sec:Tdri}

%
%%% Sub-section 2.1 %%%   %%%%%%%%%%%%%%%%%%%%%%%%%%%%%%%%%%%%%%%%%%%%%%
%
\subsection{The first topological model of an active region}
   \label{sub:Tdri}

%
%%% Sub-sub-section 2.1.1 %%%   %%%%%%%%%%%%%%%%%%%%%%%%%%%%%%%%%%%%%%%%
%
\subsubsection{The potential field approximation}

Gorbachev and Somov~[7,8]
%%% Gorbachev and Somov~(1989, 1990) %%%
have developed a 3D model for a {\em potential\/} field in the AR~2776
with an extended flare of~1980 November~5.
Before discussing the AR and the flare (see
Section~\ref{sub:SfN5}),
%
%%% Sub-section 2.2 %%%
%
let us consider, at first, the general properties of this class of
models called {\em topological\/}.

%
%%%%%% Figure 2 %%%%%%   %%%%%%%%%%%%%%%%%%%%%%%%%%%%%%%%%%%%%%%%%%%%%%%
%
\begin{figure}[htb]
   \epsfxsize=105mm
   \centerline{\epsfbox{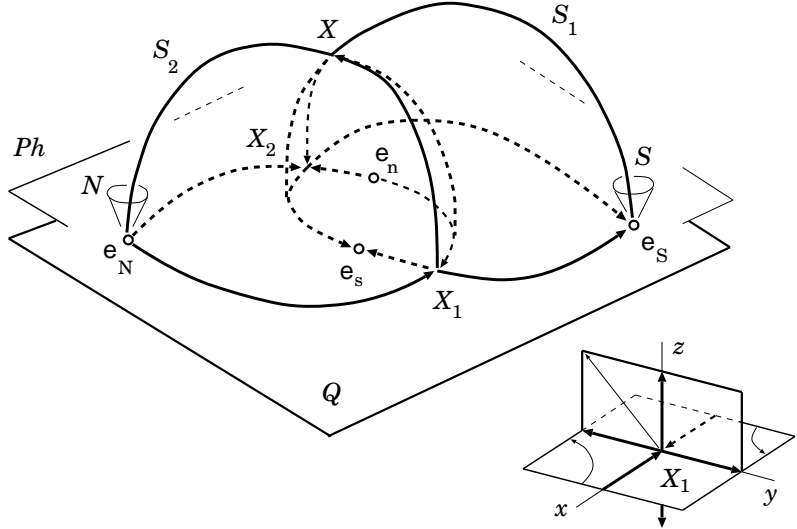}}
\caption{A model for the magnetic field of four sunspots of
         pairwise opposite polarity.
         The sunspots~$ N $ and~$ S $ in the photospheric
         plane~$ Ph $.
         The separat\-rices~$ S_{\rm 1} $ and~$ S_{\, \rm 2} $ cross
         at the separator~$ X_{\rm 1} X X_{\rm 2} $ above the
         plane~$ Q $ of the effective magnetic
         charges~$ e_{_{\rm N}} $, $ e_{_{\rm S}} $, etc. }
   \label{fig2}
\end{figure}

In the simplest model
(Gorbachev and So\-mov [9]),
%%% (Gorbachev and So\-mov, 1988), %%%
four field sources -- the magnetic
``charges''~$ e_{_{\rm N}} $ and $ e_{_{\rm S}} $, $ e_{\rm n} $
and~$ e_{\rm s} $, located in the plane~$ Q $ under the
photosphere~$ Ph $
%
%%% Figure 2 %%%%
%
(Figure~\ref{fig2}) -- are used to reproduce the main features of
the observed field in the photosphere related to the four most
important sunspots:~$ N, \, S, \, n $ and~$ s $.
As a consequence, the {\em quad\-ru\-pole\/} model reproduces only the
large-scale features of the actual field in the corona related to
these sunspots.
As a minimum, the four sources are necessarily to describe two
interacting magnetic fluxes having the two sources per each.
The larger number of sources are not necessarily much better.

The main features are two magnetic surfaces called the
{\em se\-pa\-rat\-rices\/}:~$ S_{\rm 1} $ and~$ S_{\, \rm 2} $
(Figure~\ref{fig2}).
%
%%% Figure 2 %%%
%
They divide the whole space above the plane~$ Q $ into four regions
and, correspondingly, the whole field into four magnetic fluxes having
different linkages.
The field lines are grouped into four regions according to their
termini.
The se\-pa\-rat\-rices are formed from lines beginning or
ending at magnetic zeroth points~$ X_{1} $ and $ X_{\, 2} $.
For example, the field lines originating at the point~$ X_{1} $ form
a se\-pa\-rat\-rix surface~$ S_{\rm 1} $.

The topologically singular field
line~$ X_{\rm 1} X X_{\rm 2} $, lying at the intersection of the
se\-pa\-rat\-rices, belongs to all four fluxes (two reconnecting
and two reconnected fluxes) that interact at this line, the 3D
magnetic {\em separator\/}.
So the separator separates the interacting fluxes by
the se\-pa\-rat\-rices.
Detection of a separator, as a field line that connects two zeroth
points in the Earth magneto\-tail, by the four {\em Cluster\/}
spacecraft
(Xiao et al.~[10])
%%% (Xiao et al.,~2007) %%%
provides an important step towards establishing an observational
framework of 3D reconnection.

%
%%% Sub-sub-section 2.1.2 %%%   %%%%%%%%%%%%%%%%%%%%%%%%%%%%%%%%%%%%%%%%
%
\subsubsection{Classification of zeroth points}

Let us clarify properties of the zeroth points of the magnetic field
discussed above.
In the vicinity of a zeroth point~$ {X}_{ i} $ located at
$ {\bf r} = {\bf r}_{ i} $ in the plane~$ Q $, the vector of
field is represented in the form
% (2)
\begin{equation}
    B_{\alpha} ( {\bf r} )
  = \, \frac{ \partial \psi }{ \partial r_{\alpha} }
  = M_{\alpha \beta} ( {\bf r}_{ i} ) \,
    \xi_{\beta} \, .
   \label{071113}
\end{equation}
Here $ \psi  ( {\bf r} ) $ is the potential of the field,
the vector~$ {\bf \xi} = {\bf r} - {\bf r}_{ i} $,
the index~$ i = 1,2 $\,; the Greek indices
$ \alpha, \beta = 1,2,3 $\,.
The symmetric matrix
% (3)
\begin{equation}
    M_{\alpha \beta} ( {\bf r}_{ i} )
  = \, \frac{ \partial^{2} \psi \, ( {\bf r}_{ i} ) }{
    \partial r_{\alpha} \, \partial r_{\beta} }
  = \left|
   \begin{array}{ccc}
     \lambda_{\, 1} & 0 & 0 \\
%\\
   0  &  \lambda_{\, 2}   & 0  \\
%\\
   0 & 0 &  \lambda_{\, 3}
   \end{array}
    \right|
    \,
   \label{071113a}
\end{equation}
in the frame of coordinates~$ r_{\alpha}^{\, \prime} $ related to
the eigenvectors~$ {\bf e}_{\alpha}^{\, \prime} $,
$ \lambda_{\, \alpha } $ are the eigenvalues of the matrix.
Because the field is potential, all the
$ \lambda_{\, \alpha } $ are real numbers.
Positive eigenvalues~$ \lambda_{\, \alpha } $ correspond to the
field lines emerging from a zeroth point and
negative~$ \lambda_{\, \alpha } $ correspond to the lines
arriving at a point~$ {\bf r}_{ i} $.

If the determinant
% (4)
\begin{equation}
    {\rm det} \, M_{\alpha \beta} ( {\bf r}_{ i} )
  = \lambda_{\, 1} \, \lambda_{\, 2} \, \lambda_{\, 3}
    \neq 0 \, ,
    \label{071113b}
\end{equation}
then the zeroth point is called {\em non-degenerate\/}.
Such a point is {\em isolated\/},
i.e., the magnetic field in its vicinity does not vanish.
Since $ {\rm div} \: {\bf B} = 0 $, we obtain a condition
% (5)
\begin{equation}
    \lambda_{\, 1} + \lambda_{\, 2} + \lambda_{\, 3} = 0 \, .
    \label{071113c}
\end{equation}
Therefore there exist only three following classes of zeroth points.

\vspace{0.8mm}

{\bf Type A}. All eigenvalues~$ \lambda_{\, \alpha } \neq 0 $.
This corresponds to two zeroth points shown in
Figure~\ref{fig2}.
%
%%% Figure 2 %%%
%
At the first one~$ X_{\, 1} $
% (6)
\begin{equation}
    \lambda_{\, 1} < 0\, , \quad
    \lambda_{\, 2} > 0\, , \quad
    \lambda_{\, 3} > 0\, .
    \label{071113d}
\end{equation}
Using the analogy with fluid flow, we say that field lines
``flow in" along the $ x $ axis and ``flow out" along
the se\-pa\-rat\-rix plane~$ ( y,  z ) $ as
shown in the left bottom insert in
Figure~\ref{fig2}.
%
%%% Figure 2 %%%
%
We shall call this subclass of non-degenerate zeroth point as
the Type~A$-$ since $ \lambda_{\, 1} < 0\, $.

On the contrary, at the point~~$ X_{\, 2} $
% (7)
\begin{equation}
    \lambda_{\, 1} > 0\, , \quad
    \lambda_{\, 2} < 0\, , \quad
    \lambda_{\, 3} < 0\, .
    \label{071113e}
\end{equation}
This will be the Type~A+ since $ \lambda_{\, 1} > 0\, $.

\vspace{0.8mm}

{\bf Type B}. This is a degenerate case
% (8)
\begin{equation}
    \lambda_{\, 1} = - \lambda_{\, 2} \neq 0\, , \quad
    \lambda_{\, 3} = 0\, .
    \label{071113f}
\end{equation}
It can occur, for example, if a zeroth point belongs to a
{\em neutral line\/}, i.e. a line consisting of zeroth points of the
hyperbolic type.

\vspace{0.8mm}

{\bf Type C}. This is another degenerate case
% (9)
\begin{equation}
    \lambda_{\, 1} = \lambda_{\, 2} = \lambda_{\, 3} = 0\, .
    \label{071113g}
\end{equation}
It means that three neutral lines intersect at the
point~$ {\bf r}_{ i} $\,.

%
%%% Sub-sub-section 2.1.3 %%%   %%%%%%%%%%%%%%%%%%%%%%%%%%%%%%%%%%%%%%%%
%
\subsubsection{The number of zeroth points}

In order to use some general theorems of differential geometry, we have
to smooth the point charges of magnetic field.
The smoothed potential~$ \psi $ of a positive source has a maximum at
the point where this charge is located.
Hence at this point all eigenvalues~$ \lambda_{\, \alpha} > 0 $.
At a singular point with a negative charge,
all~$ \lambda_{\, \alpha} < 0 $, and the potential has a
minimum.

In order to establish a relation between the number of
non-degenerate points, let us introduce the {\em topological index\/}
% (10)
\begin{equation}
    I_{top} \hspace{0.2mm} ( {\bf r}_{ i} )
  = {\rm sign}
    \left[ \, {\rm det} \, M_{\alpha \beta} ( {\bf r}_{ i} ) \,
    \right]
  = {\rm sign}
    \left( \,
    \lambda_{\, 1} \, \lambda_{\, 2} \, \lambda_{\, 3} \,
    \right) \, .
    \label{Itop}
\end{equation}
Here the function $ {\rm sign} \, (x) = + 1 $ if $ x > 0 $ and
$ {\rm sign} \, (x) = - 1 $ if $ x < 0 $.
Thus the possible types of points have the indices presented in
Table~\ref{Table1}.
%
%%% Table 1 %%%
%

%
%%% Table 1 %%%   %%%%%%%%%%%%%%%%%%%%%%%%%%%%%%%%%%%%%%%%%%%%%%%%%%%%%%
%
\begin{table}
\caption{Topological indices of singular and zeroth points}

\vspace{1.5mm}

\begin{center}
\begin{tabular} {|l|c|}
\hline
                &             \\
\,\, Type of a point & $ \,\, I_{top} $ \, \\
                &             \\
\hline
                        &      \\
\,\, Maximum of potential~$ \psi $     & $+1$ \\
\,\, Minimum of potential~$ \psi $     & $-1$ \\
\,\, Zeroth point of Type A$+$ \,\, & $+1$  \\
\,\, Zeroth point of Type A$-$ & $-1$ \\
                          &    \\
     \hline
\end{tabular}
\end{center}
     \label{Table1}
\end{table}

\vspace{1mm}

According to
Dubrovin et al.~[11]
%%% Dubrovin et al.~(1986), %%%
(see Part~II, Ch.~3, \S~14),
with some difference in notations that are standard in physics,
we formulate the following statement.

\vspace{1mm}

\noindent
{\bf Theorem~1}: For a 3D potential magnetic field, in a general
3D case of magnetic source location,
% (11)
\begin{equation}
    \sum_{i} \, I_{top} \hspace{0.1mm} ( {\bf r}_{ i} ) =
    \frac{ 1 }{ 4 \pi }
    \int {\bf B}
    \left[ \,
    \frac{\partial  {\bf B} }{\partial \hspace{0.1mm} \theta} \,
    \times
    \frac{\partial  {\bf B} }{\partial \varphi } \,
    \right] \,
    \frac{ d \hspace{0.1mm} \theta \, d \varphi }{ B^3 } \,
    = J^{\, (3)} \, ,
    \label{Itop+}
\end{equation}
where the integral is taken over the sphere of infinite radius.

\vspace{1.5mm}

If the total magnetic charge
% (12)
\begin{equation}
    e_{\, tot} = \sum_{i} e_{i} > 0 \, ,
    \label{071124}
\end{equation}
then the integral~$ J^{\, (3)} = + 1 $.
Let $ N_{max} $ ($ N_{min} $) is the number of maxima (minima) of the
potential~$ \psi $ of magnetic field~$ {\bf B} $, $ N_{A+} $
($ N_{A-} $) is the number of zeroth points of Type A$+$ (A$-$), then
according to~(\ref{Itop+}) we have
% (13)
\begin{equation}
    N_{max} - N_{min} +  N_{A+} -  N_{A-} = + 1 \, .
    \label{071124a}
\end{equation}
If $ e_{\, tot} < 0 $ then $ J^{\, (3)} = - 1 $, and we write $ - 1 $
on the right-hand side of Equation~(\ref{071124a}).

If the total charge
% (14)
\begin{equation}
    e_{\, tot} %%% = \sum_{i} e_{i} %%%%%%%%%%%%%%%%%%%%%%%%%%%%%%%%%%%%
    = 0 \, ,
    \label{071124b}
\end{equation}
then $ J^{\, (3)} = 0 $, and the following relation is valid:
% (15)
\begin{equation}
    N_{max} - N_{min} +  N_{A+} -  N_{A-} = 0 \, .
    \label{071124c}
\end{equation}

For the magnetic field shown in
Figure~\ref{fig2},
%
%%% Figure 2 %%%
%
$ N_{max} = N_{min} = 2 $.
Hence formula~(\ref{071124c}) gives us an equation
% (16)
\begin{equation}
    N_{A+} =  N_{A-} \,
    \label{071124c+}
\end{equation}
but not the total number of zeroth points, that we need.

\vspace{0.3mm}

However, if all the charges are located in the plane~Q
%%% (the plane~$ z = 0 $ in what follows), %%%%%%%%%%%%%%%%%%%%%%%%%%%%%
as illustrated by
Figure~\ref{fig2},
%
%%% Figure 2 %%%
%
then we can obtain an additional information about the
number of zeroth points, another equation.
Let us determine the topological index for the 2D vector field in the
same plane
% (17)
\begin{equation}
    {\bf B}^{\, (2)} (x,y)
  = \left\{ \, B_{x}, B_{y} \,
    \right\} \,
    \label{071124d}
\end{equation}
following general definition from
Dubrovin et al.~[11]
%%% Dubro\-vin et al.~(1986) %%%
% (18)
\begin{equation}
    I_{top}^{\, (2)}  \hspace{0.2mm} ( {\bf r}_{ i} )
%  = {\rm sign}
%    \left[ \, {\rm det} \, M_{\alpha \beta} ( {\bf r}_{ i} ) \,
%    \right]
  = {\rm sign}
    \left( \,
    \lambda_{\, 1} \, \lambda_{\, 2} \,
    \right) \, .
    \label{071124e}
\end{equation}
Hence the possible types of non-degenerate points have the
indices presented in Table~\ref{Table2}.

%
%%% Table 2 %%%   %%%%%%%%%%%%%%%%%%%%%%%%%%%%%%%%%%%%%%%%%%%%%%%%%%%%%%
%
\begin{table}[htb]
\caption{Topological indices of singular and zeroth points of the
         2D field in the plane~$ Q $}

\vspace{1.5mm}

\begin{center}
\begin{tabular} {|l|c|}
\hline
                &             \\
\,\, Type of a point & $ \,\, I_{top}^{\, (2)}  $ \, \\
                &             \\
\hline
                        &      \\
\,\, Maximum of potential~$ \psi $ \,\, & $+1$ \\
\,\, Minimum of potential~$ \psi $      & $+1$ \\
\,\, Zeroth point (a saddle) & $-1$ \\
                          &    \\
     \hline
\end{tabular}
\end{center}
     \label{Table2}
\end{table}

In this case of a plane arrangement of magnetic charges,
we have to rewrite Theorem~1 as follows.

\vspace{1mm}

\noindent
{\bf Theorem~2}:
For a potential magnetic field in the plane~$ Q $, in which the
charges are located,
% (19)
\begin{equation}
    \sum_{i} \, I_{top}^{\, (2)} \hspace{0.1mm} ( {\bf r}_{ i} ) =
    \frac{ 1 }{ 2 \pi }
    \int
    \left[ \, {\bf B}^{\, (2)} (x,y) \, \right]^{-2} \,
    \left( \,
    B_{x} \, d B_{y} -  B_{y} \, d B_{x} \,
    \right) \,
    = J^{\, (2)}  \, .
    \label{071124f}
\end{equation}
Here the integral is taken over the circle of infinite radius in the
positive direction of circulation.

\vspace{1mm}

For a plane field~(\ref{071124d}) of the dipole type (i.e.,
$ e_{\, tot} = 0 $ and an effective dipole moment~$ {\bf m} \neq 0 $)
the integral~$ J^{\, (2)} = 2 $.
Thus we have
% (20)
\begin{equation}
    N_{max}^{\, (2)} + N_{min}^{\, (2)} -  N_{A}^{\, (2)}
    = 2 \, ,
    \label{071124g}
\end{equation}
where
$ N_{max}^{\, (2)} $ ($ N_{min}^{\, (2)} $) is the number of maxima
(minima) of the potential~$ \psi =  \psi (x, y, 0) $ in the
plane~$ z = 0 $, $ N_{A}^{\, (2)} $ is the number of zeroth points,
the saddles.
Because of the symmetry relative to the plane~$ Q $, everywhere
in the plane (outside of the charges) $ B_{z} = 0 $.
Therefore the saddles of the plane field are zeroth points of a
3D field.

For the field shown in
Figure~\ref{fig2},
%
%%% Figure 2 %%%
%
$ N_{max}^{\, (2)} = 2 $ and $ N_{min}^{\, (2)} = 2 $.
Hence, according to formula~(\ref{071124g}), the number of zeroth
points $ N_{A}^{\, (2)} = 2 $.
By using Equation~(\ref{071124c+}), we conclude that, in the simple
case under consideration,
% (21)
\begin{equation}
    N_{A+} =  N_{A-} = 1 \, .
    \quad \quad \quad \quad {\rm Q.e.d.}
    \label{071124c+X}
\end{equation}

\vspace{0.5mm}

%
%%% Sub-section 2.1.4 %%%   %%%%%%%%%%%%%%%%%%%%%%%%%%%%%%%%%%%%%%%%%%%%
%
\subsubsection{The electric currents needed}

The potential field model does not include any currents and so cannot
model an energy stored in the fields and released in flares.
Therefore here we introduce some currents and energetics to a flare
model.

The topological model under consideration does not mean, of course,
that we assume the existence of magnetic charges under the
photosphere as well as the real zeroth points~$ X_{1} $ and~$ X_{\,2} $
in the plane~$ Q $ which does not exist either.
We assume that above the photosphere the large-scale field can be
described in terms of such a simple model.
If the magnetic sources move and change, the field also changes.
It is across the separator that the magnetic fluxes are reconnected
so that the field could remain potential, if there were no plasma.

\begin{figure}[htb]
   \epsfysize=46mm
   \centerline{\epsfbox{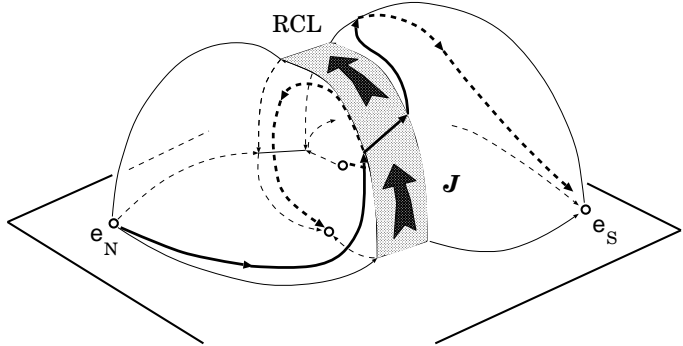}}
   \caption{A schematic view of a RCL with a total current~$ J $ at the
            separator.}
   \label{fig3}
\end{figure}

In the presence of a plasma of low resistivity, the separator plays
the same role as the hyperbolic zeroth line of magnetic field,
familiar from 2D MHD problems
(Syrovatskii~[12],
Brushlinskii et al.~[13],
Biskamp~[14,15];
%%% (Syro\-vatskii,~1962; %%%
%%% Brush\-linskii et al.,~1980; %%%
%%% Biskamp,~1986, 1997); %%%
see review of a current state of numerical simulations and
laboratory experiments in
Yamada et al.~[16].
%%% Yamada et al.~(2007). %%%
In particular, as soon as the separator appears, the electric
field~$ {\bf E}_{\, 0} $ induced by the varying magnetic field
produces an electric current~$ {\bf J} $ along the separator.
The current interacts with the potential field in such a way
that the current assumes the shape of a thin wide current layer
(RCL in
Figure~\ref{fig3}).
%
%%% Figure 3 %%%
%

%
%%%%% Figure 3 %%%%%%   %%%%%%%%%%%%%%%%%%%%%%%%%%%%%%%%%%%%%%%%%%%%%%%%
%

In a low-resistivity plasma the current layer hinders the
redistribution of magnetic fluxes, their reconnection.
This results in an energy being stored in the form of magnetic energy
of a current layer -- the free energy.
Therefore the slowly-reconnecting current layer appears at the
separator
(Syro\-vatskii~[17],
Long\-cope and Cow\-ley~[18])
%%% (Syro\-vatskii,~1981; %%%
%%% Long\-cope and Cow\-ley,~1996) %%%
in a pre-flare stage.
If for some reason (see
Somov~[19],
Cassak et al.~[20],
Uzdensky~[21])
%%% Somov,~2006; %%%
%%% Cassak et al.,~2007; %%%
%%% Uzdensky,~2007) %%%
reconnection becomes fast, the free magnetic energy is rapidly
converted into energy of particles.
This is a {\em flare\/}.
The rapidly-reconnecting current layer,
being in a high-temperature turbulent-current state
(Section~\ref{sec:BPoH}),
%
%%% Section 4 %%%
%
provides the energy fluxes along the reconnected field lines.

Reconnection for 3D models without zeroth points in and above the
photosphere, as we assumed in
Figure~\ref{fig3},
%
%%% Figure 3 %%%
%
is qualitatively the same as the 2D reconnection in the models with a
non-zero {\em longitudinal\/} magnetic
field~$ {\bf B}_{\, \parallel} \, $.
The inductive electric field along the separator is the driving
force of reconnection.
The field lines going to reconnect are pushed into the RCL and
the lines coming out of the RCL are pulled
by the longitudinal electric field.
Thus reconnection in 3D should be taken to mean effects associated
with the RCL, as in 2D
(Greene~[22],
Lau and Finn~[23]).
%%% (Greene,~1988; %%%
%%% Lau and Finn,~1990). %%%

An actual 3D reconnection at a separator proceeds in the presence
of an increasing (or decreasing) longitudinal field
$ {\bf B}_{\, \parallel} \, $.
What factors do determine the increase (or decrease) of this field? -
The first of them is the global field configuration, i.e. the
relative position of the magnetic field sources in an AR.
It determines the position of a separator and the value of the
longitudinal field at the separator and in its vicinity.
This field is not uniform, of course.

The second factor is evolution of the global configuration, more
exactly, the electric field~$ {\bf E}_{\, 0} $
related to the evolution and responsible for driven reconnection at
a separator.
The direction of reconnection - with an increase (or decrease) of
the longitudinal magnetic field - depends on the sign of the
electric field projection on the separator, i.e. on the sign of the
scalar product
$ ( \, {\bf E}_{\, 0} \cdot {\bf B}_{\, \parallel} \, ) $.
In general, this sign can be plus or minus with equal probabilities,
if there are no preferential configurations of the global field or no
preferential directions of the AR evolution.

%
%%%%%%%%%%%%%%%%%%%%%%%%%%%%%%%%%%%%%%%%%%%%%%%%%%%%%%%%%%%%%%%%%%%%%%%%
%                                                                      %
%%% Section 2.2 %%%   %%%%%%%%%%%%%%%%%%%%%%%%%%%%%%%%%%%%%%%%%%%%%%%%%%
%
\subsection{The solar flare on 5 November 1980} %%%%%%%%%%%%%%%%%%%%%%%%
   \label{sub:SfN5}

%
%%% Sub-section 2.2.1 %%%   %%%%%%%%%%%%%%%%%%%%%%%%%%%%%%%%%%%%%%%%%%%%
%
\subsubsection{Observed and model magnetograms}

\begin{figure}[htb]
   \epsfysize=54mm
   \centerline{\epsfbox{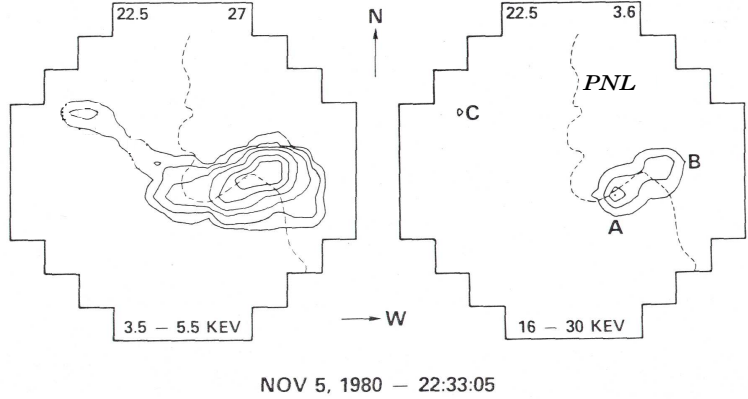}}
   \caption{The flare of~1980 November~5 as observed by
            {\em SMM\/}. }
   \label{fig4}
\end{figure}

The first well-studied example was the extended 1B/M4~flare on~5
November 1980 (e.g. Rust and So\-mov~[24]) observed by the
satellite~{\em SMM\/}
(Figure~\ref{fig4}).
%
%%% Figure 4 %%%
%
%
%%%%%% Figure 4 %%%%%%   %%%%%%%%%%%%%%%%%%%%%%%%%%%%%%%%%%%%%%%%%%%%%%%
%
Three bright {\em hard\/} X-ray (HXR) kernels~A, B and C are well
dis\-tin\-guished in the right panel.
The {\em soft\/} X-ray (SXR) elements of the flare, as seen in the
left panel, consist presumably of two overlapping coronal loops,~AB
and~BC.
Their foot\-points in the chromosphere have been identified as
the sources of HXR and H$ \alpha $ emission.

%
%%%%%% Figure 5 %%%%%%   %%%%%%%%%%%%%%%%%%%%%%%%%%%%%%%%%%%%%%%%%%%%%%%
%
\begin{figure}[htb]
   \epsfysize=79mm
   \centerline{\epsfbox{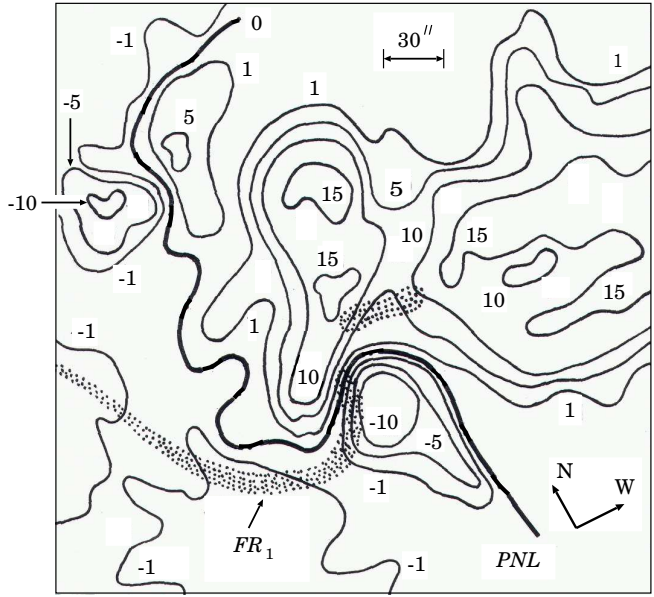}}
   \caption{The line-of-sight component of magnetic field
            in AR~2776.
            Two flare ribbons observed in H$ \alpha $ line
            are shown as shadow areas;
            $ FR_{1} $ is the longer one.}
   \label{fig5}
\end{figure}

Figure~\ref{fig5}
%
%%% Figure 5 %%%
%
shows the line-of-sight magneto\-gram of the AR~2776 where the flare
occurred.
Numbers shows a magnitude of the field measured in
$ 10^2 $~G.
Two narrow flare ribbons are shown on either side of the
photospheric neutral line~($ PNL $).
Let us identify the four largest regions in which the field
of a single polarity is concentrated:
two of northern polarity and two of souther polarity.
Since the AR is comparatively close to the center of the solar disk,
the magneto\-gram represents the vertical component~$ B_{z} $ at the
photospheric level fairly well.

%
%%%%%%  Figure 6 %%%%%%   %%%%%%%%%%%%%%%%%%%%%%%%%%%%%%%%%%%%%%%%%%%%%%
%
\begin{figure}
   \epsfxsize=152mm
   \centerline{\epsfbox{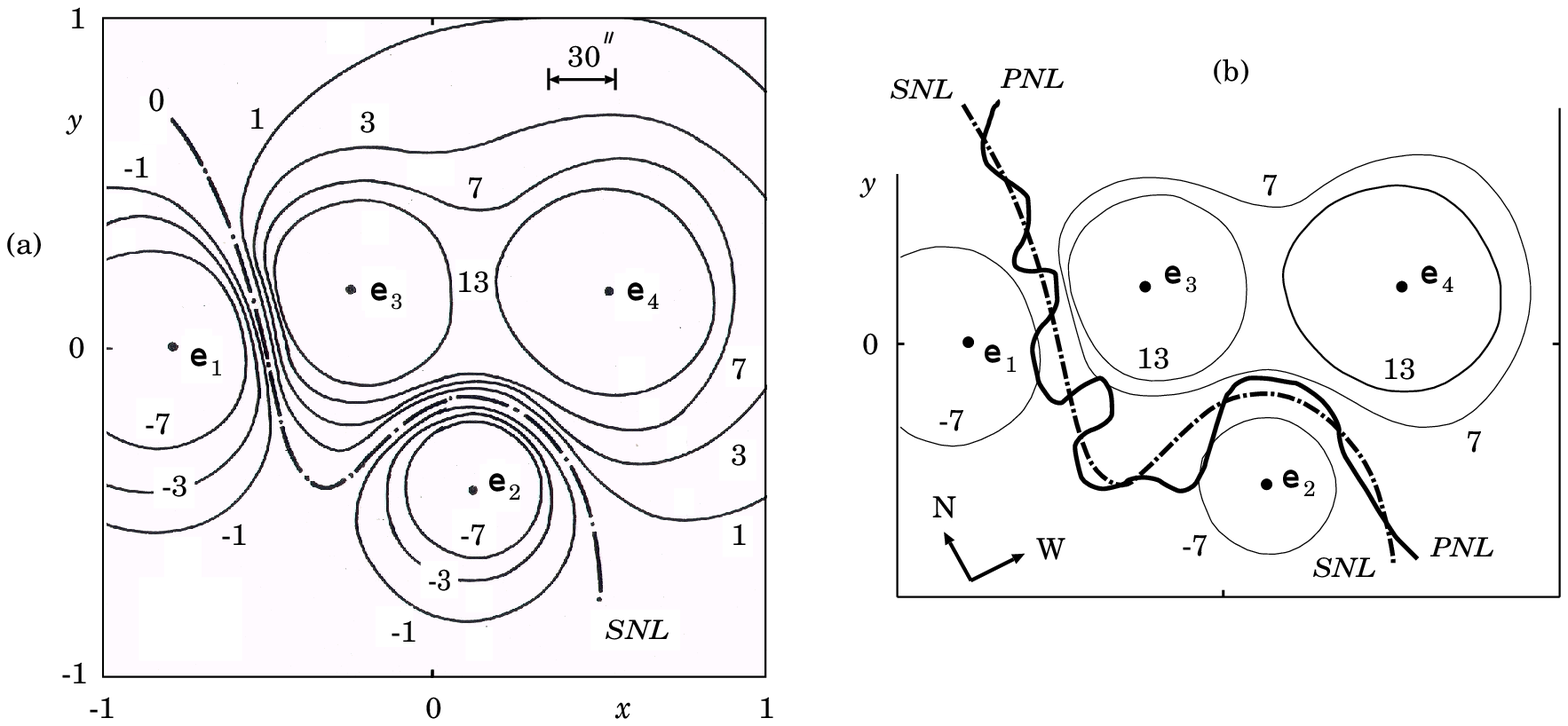}}
   \caption{The model magneto\-gram.
            (a) Positions and magnitudes of the sources fit the main
            features of the
            observed magneto\-gram.
            (b) The calculated {\em simplified\/} neutral
            line~($ SNL $) approximates the observed
            photospheric neutral line~($ PNL $). }
   \label{fig6}
\end{figure}

Let us make a model of the magneto\-gram shown in
Figure~\ref{fig5}.
%
%%% Figure 5 %%%
%
With this aim, we calculate the vertical component~$ B_{z} $ of
magnetic field in the photospheric plane~$ Ph$ (plane~$ z = 0 $ in
this Section) in the potential approximation by using the simple
formula
% (22)
\begin{equation}
 {\bf B} \, (x,y,z)
    = \sum \limits_{i=1}^{4}
    \frac{ e_i }{ | \, {\bf r} - {\bf r}_i \, |^{\, 2} } % \cdot
    \times
    \frac{ {\bf r} - {\bf r}_{i} }{ | \, {\bf r}-{\bf r}_{i} \, |}
    \, .
    \label{070806}
\end{equation}
Here $ e_{i} $ are the effective charges of field, and
$ {\bf r}_{i} $ are their radius vectors:
$$
\begin{array}[]{ll}
  e_{1} = - \, 2.5 \, , \quad &
  {\bf r}_{1} = \{ \, - \, 0.8 \, , 0\, \, , - \, 0.1 \, \} \, ; \\
  e_{2} = - \, 1.0 \, , &
  {\bf r}_{2} = \{ \, 0.1 \, , - \, 0.4 \, , - \, 0.1 \, \} \, ; \\
  e_{3} = + \, 3.0 \, , &
  {\bf r}_{3} = \{ \, - \, 0.25 \, , 0.15 \, , - \, 0.1 \, \} \, ; \\
  e_{4} = + \, 4.5 \, , &
  {\bf r}_{4} = \{ \, 0.5 \, , 0.16 \, , - \, 0.1 \, \} \, .\\
\end{array}
$$
Note that the total magnetic charge
% (23)
\begin{equation}
    e_{\, tot} = %%% \sum_{i} e_{i} =
    4 > 0 \, .
    \label{071124n}
\end{equation}
Thus, in order to determine a relation between the number of
zeroth points and the numbers of field sources, we have
to use formula~(\ref{071124a}).
It gives us an equation
% (24)
\begin{equation}
    N_{A+} =  N_{A-} + 1 \, .
    \label{071124an}
\end{equation}
We shall come back to this equation, considering a topological
portrait of the AR.

The parameters~$ e_i $ and~$ {\bf r}_{i} $ have been selected in
order to reproduce in an optimal way the fluxes of
individual polarities in the photospheric plane
(Figure~\ref{fig6}a)
%
%%% Figure 6 %%%
%
as well the shape of the photospheric neutral line
($ PNL $ in Figure~\ref{fig6}b).
%
%%% Figure 6 %%%
%
The calculated {\em simplified\/} neutral line~($ SNL $) smoothes
the curve~$ PNL $ in small scales but conserves its $ S $-type
shape in a large scale, in the linear scale of the AR.
In Figure~\ref{fig6},
%
%%% Figure 6 %%%
%
the length unit equals
150$^{\, \prime \prime} \approx 1.1 \times 10^{10} $~cm;
the numbers near the curves show the values of the vertical component
of the field measured in 10$ ^{2} $~G.

Because the simplest model uses a minimal number of
magnetic sources -- four, which is necessary to describe the minimal
number of interacting fluxes -- two, we call it the
{\em quad\-ru\-pole-type\/} model
(Figure~\ref{fig6}).
%
%%% Figure 6 %%%
%
This is not an exact definition (because in the case under
consideration
$ e_{_{\rm N}} \not= - \, e_{_{\rm S}} $ and
$ e_{\rm n} \not= - \, e_{\rm s} $) but it is convenient for
people who know the {\em exact-quad\-ru\-pole\/} model
(Sweet~[25]).
%%% (Sweet,~1969). %%%
In fact, the difference, the presence of another separator in the
model under consideration, is not small and can be
significant for actual ARs.
The second separator may be important to give accelerated
particles a way to escape out of an AR in the interplanetary space.

We calculate the field lines integrating the ordinary
differential equations
% (25)
\begin{equation}
    \frac{ d x}{B_x}
  = \frac{ d y}{B_y}
  = \frac{ d z}{B_z}\, .
    \label{070806a}
\end{equation}
Here $ ( dx, dy, dz ) $ is an arch element directed along a field line.
The vector~$ {\bf B} $ is determined at each point by
formula~(\ref{070806}).

%
%%% Sub-section 2.2.2 %%%   %%%%%%%%%%%%%%%%%%%%%%%%%%%%%%%%%%%%%%%%%%%%
%
\subsubsection{Topological portrait of the active region} %%%%%%%%%%%%%%
\begin{figure}[htb]
   \epsfysize=79mm
   \centerline{\epsfbox{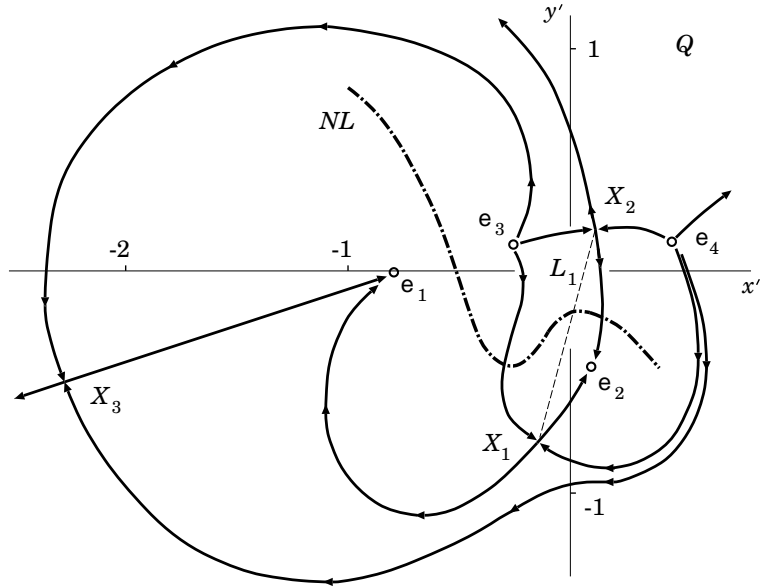}}
   \caption{The topological portrait of the AR 2776 where the
            extended 1B/M4 solar flare
            of 1980 November~5 occurred.}
   \label{fig7}
\end{figure}

   \label{sub:SfN5+}
The topological properties of magnetic field are determined by
the number and locations of zeroth points.
In the case under consideration, there are three points
in the plane~$ Q $ (here $ z = - 0.1 $) shown in
Figure~\ref{fig7}.
%
%%% Figure 7 %%%
%
The coordinates of these points are:
$$
\begin{array}[]{ll}
  {X}_{1} = \{ \, - \, 0.133 \, , - \, 0.739 \, , - \, 0.1 \, \} \, , \\
  {X}_{2} = \{ \, 0.076 \, ,  \, 0.179 \, , - \, 0.1 \, \} \, , \\
  {X}_{3} = \{ \, - \, 2.26 \, , - \, 0.439 \, , - \, 0.1 \, \} \, .\\
\end{array}
$$

%
%%%%%% Figure 7 %%%%%%   %%%%%%%%%%%%%%%%%%%%%%%%%%%%%%%%%%%%%%%%%%%%%%%
%

%
%%% Figure 7 %%%
%
Figure~\ref{fig7} shows the topologically important field lines
in the plane~($ x^{\prime}, y^{\prime} $) which is the plane~$ Q $ of
the sources~$ {\rm e}_{1} $,
$ {\rm e}_{2} $, $ {\rm e}_{3} $, and $ {\rm e}_{4} $.
Since the total charge~$ e_{\, tot} > 0 $, instead of
Equation~(\ref{071124g}) for the number of zeroth points of
2D field~(\ref{071124d}), we have an equation
% (26)
\begin{equation}
    N_{max}^{\, (2)} + N_{min}^{\, (2)} -  N_{A}^{\, (2)}
    = 1 \, ,
    \label{071124gnX}
\end{equation}
because according to formula~(\ref{071124f}) we find the
value~$ J^{\, (2)} = 1 $.
It follows from (\ref{071124gnX}) that the total number of
zeroth points
% (27)
\begin{equation}
    N_{A}^{\, (2)}
    = 3 \, .
    \label{071130a}
\end{equation}
On the other hand,
% (28)
\begin{equation}
    N_{A}^{\, (2)}
    = N_{A+} +  N_{A-} \, .
    \label{071130b}
\end{equation}
With account of (\ref{071124an}) taken, we conclude that
$ N_{A+} = 2 $ and $ N_{A-} = 1 $.

The field lines shown in
Figure~\ref{fig7}
%
%%% Figure 7 %%%
%
play the role of separat\-rices and show the presence of
{\bf two separators\/}.
Two points~$ X_{1} $ (Type~$ A_{+} $) and $ X_{\, 2} $ (Type~$ A_{-} $)
are located in the vicinity of the magnetic sources and are connected by
the first separator shown by its projection, the thin dashed line~$ L_{1} $.
Near this main separator, the field  and its gradient are strong and
determine the flare activity of the AR.

Another separator starts from the point~$ X_{3} $
(the Type~$ A_{+} $) far away from the sources and goes much higher
above the AR to infinity.
The second separator can be responsible for flares in weaker
magnetic fields and smaller gradients high in the corona.
The second separator is also a good place for the particles accelerated
along the main separator to escape from the AR in the interplanetary
space.

\begin{figure} % [htb]
   \epsfysize=48mm
   \centerline{\epsfbox{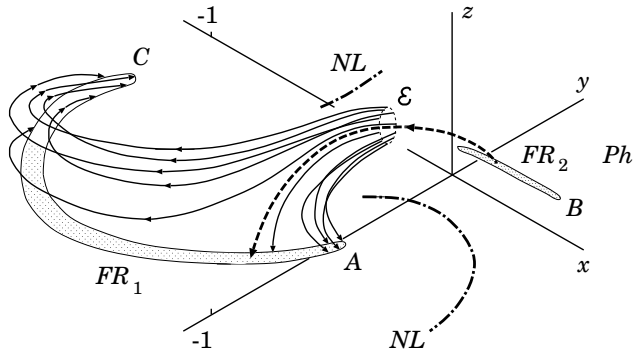}}
   \caption{Potential field lines cross a
            region of primary energy release~$ {\cal E} $
            situated at the apex of the main separator
            (boldface dashed curve).
            Two flare ribbons are formed where these lines
            cross the photospheric plane~$ Ph $.}
   \label{fig8}
\end{figure}

We suppose that a part of the flare energy is released in some
region~$ {\cal E} $ near the apex of the main separator.
The energy fluxes~$ F_{_{\cal E}} $ propagate {\em along\/}
the field lines connecting the energy source with the photosphere.
Projections of the energy source~$ {\cal E} $ on the
photospheric plane~$ Ph $ along the field lines are shown as two
``flare ribbons''~$ FR_{1} $ and~$ FR_{2}  $ in
Figure~\ref{fig8}.
%
%%% Figure 8 %%%
%
Therefore the model allows us to identify the flare bright\-enings, in
the H$ \alpha $~line as well as in EUV and HXRs, with the ribbons located
at the intersection of the separat\-rices with the chromosphere which is
placed slightly above the photospheric plane~($ Ph $).

%
%%%%%% Figure 8 %%%%%%   %%%%%%%%%%%%%%%%%%%%%%%%%%%%%%%%%%%%%%%%%%%%%%
%

The characteristic {\em saddle\/} structure of the orthogonal
(perpendicular to a separator) field~$ {\bf B}_{\bot} $ in the
vicinity of the separator leads to a spatial redistribution of the
energy flux~$ F_{_{\cal E}} $ of heat and accelerated particles.
This flux is split apart in such a way that it creates
the long-narrow ribbons.
Thus, at first, the model had reproduced the
observed features of the real solar flare.
In particular, the model predicts the simultaneous flaring of the
two chromospheric ribbons.
Moreover it predicts that a concentration of the field lines that
bring energy into the ribbons is higher at
the edges of the ribbons, i.e. at relatively compact regions
indicated as~$ A $, $ B $, and~$ C $.
Here the H$ \alpha $ bright\-enings must be especially bright.
This is consistent
with observations of H$ \alpha $ ``kernels'' in this flare.

%
%%% Section 2.3 %%%   %%%%%%%%%%%%%%%%%%%%%%%%%%%%%%%%%%%%%%%%%%%%%%%%%%
%
\subsection{Features of the flare topological model} %%%%%%%%%%%%%%%%%%%
   \label{sub:foftm}

Since in the kernels the energy fluxes are concentrated, the impulsive
heating of the chromosphere creates a fast expansion of
high-temperature plasma upwards into the corona
(e.g. So\-mov~[26]).
%%% (e.g., So\-mov,~1992). %%%
This effect is known as the chromospheric ``{\em evaporation\/}''
observed in the EUV and SXR emission.
Evaporation lights up the coronal loops in flares.

\begin{figure}[htb]
   \epsfysize=48mm
   \centerline{\epsfbox{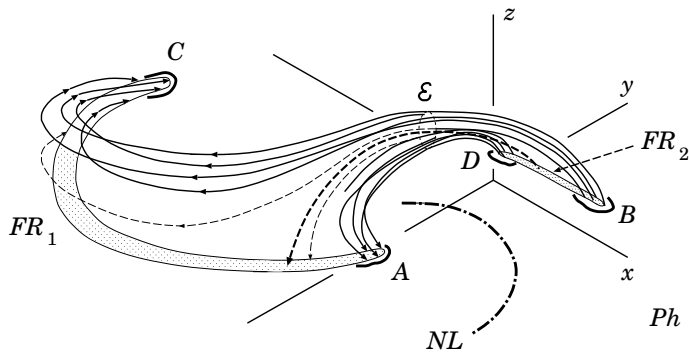}}
   \caption{Field lines that connect the H$ \alpha $~kernels~$ A $,
            $ B $, $ C $, and~$ D $.
            Chromospheric evaporation creates a picture of the
            crossing SXR loops.}
   \label{fig9}
\end{figure}

The topological model also shows that the flare ribbons as well as
their edges with H$ \alpha $~kernels are connected to a common source
of energy at the separator (see $ {\cal E} $ in
Figure~\ref{fig9}).
%
%%% Figure 9 %%%
%
Through this region all four kernels are magnetically connected to one
another.
Therefore the SXR loops look like they are crossing or touching each
other.

%
%%%%% Figure 9 %%%%%%   %%%%%%%%%%%%%%%%%%%%%%%%%%%%%%%%%%%%%%%%%%%%%%%%
%

So the reconnecting magnetic fluxes are distributed in the corona in
such a way that the {\bf two SXR loops} may look like that they
{\bf interact with each other\/}.
That is why such structures are usually considered as evidence in
favor of a model of two interacting loops
(Sakai and de Jager~[27]).
%%% (Sakai and de Jager,~1996). %%%
The difference, however, exists in the primary source of energy.
High concentrations of electric currents and twisted magnetic fields
inside the interacting loops are created by some under-photospheric
dynamo mechanism.
If these currents are mostly parallel they attract each other giving
an energy to a flare
(Gold and Hoyle~[28]).
%%% (Gold and Hoyle,~1960). %%%
On the contrary, according to the topological model, the flare energy
comes from an interaction of magnetic fluxes that can be mostly
potential.

The {\bf S-shaped structures} observed in SXRs are usually
interpreted in fa\-vour of non-potential fields.
In general, the shapes of loops are signature of the
{\em helicity\/} of their magnetic fields.
The S-shaped loops match flux tubes of {\em positive\/} helicity, and
inverse S-shaped loops match flux tubes of {\em negative\/} helicity
(Pev\-tsov et al.~[29]).
%%% (Pev\-tsov et al.,~1996). %%%
However, in
Figure~\ref{fig9},
%
%%% Figure 9 %%%
%
the S-shaped structure~$ C{\cal E}B $ connecting the bright
points~$ C $ and~$ B $ results from the computation of the potential
field.

In the AR~2776,
Den and Somov~[30]
%%% Den and Somov~(1989) %%%
found a considerable {\bf shear of a potential field\/} above the
$ PNL $ near the brightest loop~$ AB $.
Many authors concluded that an initial energy of flares is stored in
magnetic fields with large shear.
However, such flares presumably were not the case of potential field
having a minimum energy.
This means that the presence of observed shear is not a sufficient
condition for generation of a flare.

The topological model by
Gorbachev and Somov~[7]
%%%
postulated a global topology for an AR consisting of four main fluxes.
Reconnection between, for example, the upper and lower fluxes
transfers a part of the magnetic flux to the two side systems.
Antiochos~[31]
%%% Antiochos~(1998) %%%
addresses the following question: What is the minimum complexity
needed in the magnetic field of an AR so that a similar
process can occur in a fully 3D geometry?
He starts with a highly sheared field near the $ PNL $ held down by an
overlying un\-sheared field.
Antio\-chos concludes that a real AR can have much more
complexity than very simple configurations.
We expect that the large-scale topology of four-flux systems meeting
along a separator is the basic topology underlying eruptive
activity of the Sun.

On the other hand, configurations with more separators (as well as
with multiply connected magnetic sources and zeroth-point pairs; see
Parnell~[32])
%%% Parnell,~2007) %%%
have more opportunity to reconnect and would thus more likely
to produce flares.
Such complicated configuration would presumably produce many small
flares to release a large excess of energy in an AR rather
than one large flare.

%
%%% Section 3 %%%   %%%%%%%%%%%%%%%%%%%%%%%%%%%%%%%%%%%%%%%%%%%%%%%%%%%%
%
\section{Topological trigger for solar flares}
    \label{sec:Tt}

%
%%% Sub-section 3.1 %%%%%%%%%%%%%%%%%%%%%%%%%%%%%%%%%%%%%%%%%%%%%%%%%%%%
%
\subsection{What is that?}
    \label{sub:Tt}

The effect of topological trigger was suggested by
Gorbachev et al.~[33]
%%% Gorbachev et al.~(1988) %%%
in spite of a large language barrier between its mathematical
background
(Dub\-ro\-vin et al.~[11])
%%% (Dub\-ro\-vin et al.,~1986) %%%
and the physical terms used by models of
ARs.
Many people simply understood the simple topological model
but, unfortunately, not the topological trigger.

Fortunately,
Barnes~[34]
%%% Barnes~(2007) %%%
investigated a relationship between solar eruptive events and the
existence of the coronal zeroth points of field, using
a collection of over 1800 vector magneto\-grams.
Each of them was subjected to the charge topology
analysis, including determining the presence of coronal zeroth
points.
It appears that the majority of events originate in ARs above
which no zeroth point (a magnetic null) was found.
However a much larger fraction of ARs, for which a coronal
zeroth point was found, were the source of an eruption than ARs for
which no zeroth point was found.
Clearly the presence of a coronal zeroth point is an indication that
an AR is more likely to produce an eruption,
as 35~\% of the ARs for which such a point was found produced
eruptions, compared with only 13~\% of ARs for which no zeroth
point was found.
We consider this fact as indication that the topological trigger
can play a significant role in the origin of eruptive flares.
This is also consistent with the study of
Ugarte-Urra et al.~[35].
%%% Ugarte-Urra et al.~(2007). %%%

\begin{figure} [ht]
\epsfysize=38mm
   \centerline{\epsfbox{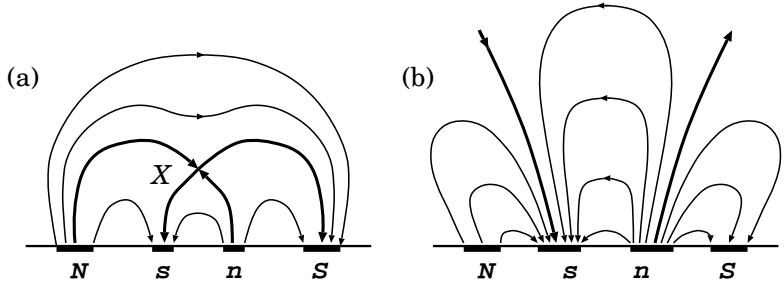}}
   \caption{Topological change of a potential field created by four
            magnetic sources.
            (a) The closed configuration with the zeroth point~$ X $.
            (b) The critical instant when the magnetic field opens. }
   \label{fig10}
\end{figure}

The possibility of a topological approach to the question of the
trigger for flares was not often discussed in the literature.
Syrovatskii and Somov~[36]
%%% Syrovat\-skii and Somov~(1980) %%%
considered a slow evolution of coronal fields and showed that, during
such an evolution, some critical state can be reached, and fast dynamic
phase of evolution begins and is accompanied by a rapid change of
magnetic topology.
For example, it is possible a rapid `{\em break-out\/}' of a `new'
magnetic flux through the `old' coronal field of an AR
(Figure~\ref{fig10}).
%
%%% Figure 10 %%%
%
%
%%%%%% Figure 10 %%%%%%   %%%%%%%%%%%%%%%%%%%%%%%%%%%%%%%%%%%%%%%%%%%%%%
%
When an effective magnetic moment of the internal growing group of
sunspots becomes nearly equal an effective moment of the AR
(Syrovatskii~[37]),
%%% (Syrovatskii,~1982), %%%
the closed configuration {\em quickly turns\/} into the open one.

In this paper we discuss another possibility.
Near a separator the longitudinal
component~$ {B}_{\, \parallel} $ dominates because the
orthogonal field~$ {\bf B}_{\bot} $ vanishes at the separator.
Reconnection in the RCL at the separator just
conserves the flux of the longitudinal field
(see Somov~[19]).
%%% (see Somov,~2006). %%%
At the separator, the orthogonal components are reconnected.
Therefore they actively participate in the connectivity change, but
the longitudinal field does not.
Thus it seems that the longitudinal field plays a passive role in the
topological aspect of the process but it influences the physical
properties of the RCL, in particular the reconnection rate.
The longitudinal field decreases compressibility of plasma flowing
into the RCL.
When the longitudinal field vanishes, the plasma becomes ``strongly
compressible'', and the RCL collapses, i.e. its width
decreases substantially.
As a result, the reconnection rate increases quickly.
However this is not the whole story.

%
%%%%%% Figure 11 %%%%%%   %%%%%%%%%%%%%%%%%%%%%%%%%%%%%%%%%%%%%%%%%%%%%%
%
\begin{figure} % [bht]
\epsfysize=61mm
   \centerline{\epsfbox{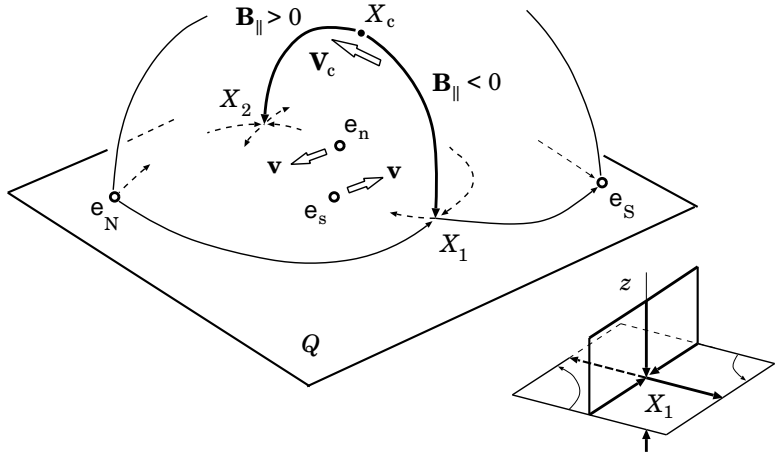}}
   \caption{The zeroth point $ X_{\, \rm c} $ rapidly moves
            along the separator and switches back the longitudinal
            component~$ {\bf B}_{\parallel} $ of magnetic field. }
   \label{fig11}
\end{figure}

The important exception constitutes a zeroth point which can appear
on the separator above the photosphere.
Gorbachev et al.~[33]
%%% Gorbachev et al.~(1988) %%%
showed that, in this case, even very slow changes in the configuration
of field sources in the photosphere can lead to a rapid
migration of such a point along the separator
(Figure~\ref{fig11})
%
%%% Figure 11 %%%
%
and to a topological trigger of a flare.
This essentially 3D effect will be considered below in more detail.
Note that the topological trigger effect is {\em not\/} a resistive
instability which leads to a change of the topology of the field
configuration from pre- to post reconnection state.
On the contrary, the topological trigger is a quick change of the
global topology, which dictates the fast reconnection of collisional
or collision\-less origin.
Thus the topological trigger is the most appropriate nomenclature to
emphasize the basic nature of the topological effect involved, and
it is a welcome usage.

%
%%%%%%%%%%%%%%%%%%%%%%%%%%%%%%%%%%%%%%%%%%%%%%%%%%%%%%%%%%%%%%%%%%%%%%%%
%
%%% Sub-section 3.2 %%%%%%%%%%%%%%%%%%%%%%%%%%%%%%%%%%%%%%%%%%%%%%%%%%%%
%
\subsection{How does the topological trigger work?}
    \label{sub:Hdiw}

Let us trace how a rapid rearrangement of the field topology occurs
under conditions of slow evolution of photospheric sources
with the total charge equals zero.
We shall arbitrary fix the positions of three charges,
while we move the fourth one along an arbitrary trajectory in the
plane~$ Q $.

Let an initial position of the moving charge corresponds to the
values of the topological indices $ I_{top} = - 1 $ and
%%% $ I_{top} = + 1 $ %%%
+ 1 for the zeroth points~$ X_{1} $ (Type~A$ - $
with $ \lambda_{z} > 0 $) and $ X_{\,2} $ (Type~A$ + $ with
$ \lambda_{z} < 0 $) respectively
(Figure~\ref{fig2}).
%
%%% Figure 2 %%%
%
Thus two points in the plane~$ Q $ have different indices.
Equation~(\ref{071124c}) is satisfied.
The separator is the field line connecting these points
without a coronal null.
This field line emerges from the point~$ X_{1} $ and is directed
along the separator to the point~$ X_{\, 2} $.
The value and sign of the magnetic-potential difference between the
zeroth points are determined from the relation
% (29)
\begin{equation}
    \psi_{\, 2} - \psi_{1}
  = \int \limits_{ 1 }^{ 2 } {\bf B} \, d \, {\bf l} \, ,
    \label{080109}
\end{equation}
where the integral is taken along the separator.
$ \psi_{\, 2} - \psi_{1} < 0 $ for the field shown in
Figure~\ref{fig2}.
%
%%% Figure 2 %%%
%

\begin{figure} [tbh]
\epsfysize=73mm
   \centerline{\epsfbox{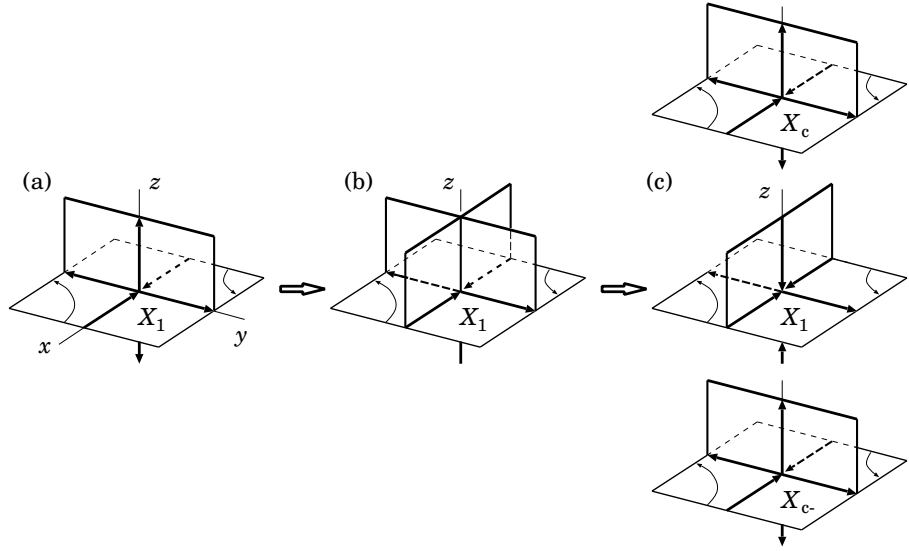}}
   \caption{Changes of the field pattern at the zeroth
            point~$ X_{1} $.
   (a) An initial state is the non-degenerate point of the Type~A$ - $
       with $ \lambda_{z} > 0 $.
   (b) A {\em degenerate\/} hyperbolic point (line) with
       $ \lambda_{z} = 0 $
       at the beginning of trigger.
   (c) After the beginning of trigger, the pattern of field is
       the non-degenerate point of the Type~A$ + $
       with $ \lambda_{z} < 0 $ and two zeroth points outside
       the plane~$ z = 0 $. }
   \label{fig12}
\end{figure}

According to
Gorbachev et al.~[33],
%%% Gorbachev et al.~(1988), %%%
the moving charge can arrive in
a narrow region (let us call it the region~$ TT $) such that both
points in the plane~$ Q $ will have the same indices
(Figure~\ref{fig11}).
%
%%% Figure 11 %%%
%
It follows from the general 3D Equation~(\ref{071124c}) that in this
case there must also exist two zeroth points outside the plane.
They are arranged symmetrically relative to the plane~$ Q $ (the
plane~$ z = 0 $ in this Section).
Figure~\ref{fig12}
%
%%% Figure 12 %%%
%
illustrates how these additional points appear.

%
%%%%%% Figure 12 %%%%%%   %%%%%%%%%%%%%%%%%%%%%%%%%%%%%%%%%%%%%%%%%%%%%%
%

Before the start of trigger, the moving charge is outside of the
region~$ TT $, the index~$ I_{top} = - 1 $ and the
eigenvalue~$ \lambda_{z} > 0$ at the non-degenerate zeroth
point~$ X_{1} $ (Figure~\ref{fig12}a).
%
%%% Figure 12 (a) %%%
%
When the moving charge crosses the boundary of the region~$ TT $,
the eigenvalue~$ \lambda_{z} $ at the point~$ X_{1} $ vanishes
(Figure~\ref{fig12}b):
%
%%% Figure 12 (b) %%%
%
% (30)
\begin{equation}
    \lambda_{z} ( X_{1} ) = 0 \, .
    \label{080101}
\end{equation}
The point becomes degenerate.
At this instant, another pair of zeroth points is born from
the point~$ X_{1} $
(Figure~\ref{fig12}c).
%
%%% Figure 12 (c) %%%
%
We shall consider only one of them, the point~$ X_{c} $ in the upper
half-space~$ z > 0 $.
This  non-degenerate point travels along the separator and
merges with the point~$ X_{\, 2} $ in the plane~$ z = 0 $ when the
moving charge emerges from the region~$ TT $.
At this instant, the eigenvalue~$ \lambda_{z} $
vanishes:
% (31)
\begin{equation}
    \lambda_{z} ( X_{\, 2} ) = 0 \, .
    \label{080102}
\end{equation}
As a result of the process described, the direction of the field at
the separator is reversed with the point~$ X_{\,2} $ of the
Type~A$ - $ with $ \lambda_{z} > 0 $.
After that, the moving charge is located outside the region~$ TT $,
there are no zeroth points outside the plane~$ z = 0 $.
Thus Equations~(\ref{080101}) and~(\ref{080102}) determine the
boundaries of the topological trigger region~$ TT $.

%
%%%%%% Figure 13 %%%%%%   %%%%%%%%%%%%%%%%%%%%%%%%%%%%%%%%%%%%%%%%%%%%%%
%
%\begin{figure} [hbt]
%\epsfysize=46mm
%   \centerline{\epsfbox{fig13.eps}}
%   \caption{The topological trigger narrow dashed region is
%            restricted by two boundaries: curves~$ C_{1} $
%            and~$ C_{2} $ corresponding to Equations~(\ref{080101})
%            and~(\ref{080102}). }
%   \label{fig13}
%\end{figure}

\begin{figure}[tbh]
\hspace*{15mm}
\parbox[b]{55mm}{ % Left box %%%%%
    \epsfysize=46mm % \hspace*{1mm}
    \epsfbox{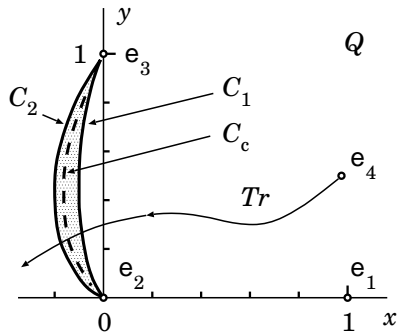}
%     \label{fig13}
    } %%%%%
    \hspace{7mm}
\parbox[b]{66mm}{ % Right box %
    \caption{The topological trigger (narrow dashed) region is
            restricted by two boundaries: curves~$ C_{1} $
            and~$ C_{2} $ corresponding to Equations~(\ref{080101})
            and~(\ref{080102}).}
    \vspace{13mm}
    \noindent %%% \label{fig13}
    }
  \label{fig13}
\end{figure}

Figure~13
%%% Figure~\ref{fig13} %%%
%
%%% Figure 13 %%%
%
presents an illustrative sample of charge configuration.
The fixed charges $ e_{1} = - e_{2} = - e_{3} = 1 $ are located in the
plane~$ Q $ at the points~(1; 0), (0; 0) and~(0; 1), respectively.
%%%;
%%%for more general case see
%%%Gorbachev et al.~(1988).
The charge~$ e_{4} = 1 $ moves along a trajectory shown by a thin
curve~$ Tr $.
In the beginning of trigger, it crosses the boundary
curve~$ C_{1} $.
Between the boundaries~$ C_{1} $ and~$ C_{2} $ there exists a
curve~$ C_{c\,} $ at which
$ \psi_{2} - \psi_{1} = 0 $.
The field at the separator changes sign at the point~$ X_{c} \,$.
Hence the sections~$ X_{1} X_{c} $ and~$ X_{c\,} X_{\, 2} $ of the
separator make contributions of opposite signs to the
integral~(\ref{080109}).
These contributions exactly compensate each other for a position
of the point~$ X_{c\,} $ when the trajectory~$ Tr $ crosses the
curve~$ C_{c\,} $.

Typically the region~$ TT $ is narrow.
That is why small shifts of the moving charge within this region lead
to large shifts of the zeroth point~$ X_{c} $ along the separator
above the plane~$ z = 0 $ just creating a {\em global bifurcation\/}.
Using the analogy with hydrodynamics, we see that the separat\-rix
plane~($ y, z $) in
Figure~\ref{fig12}a,
%
%%% Figure 12 (a) %%%
%
which plays the role of a ``hard wall'' for
``flowing in'' magnetic flux, is quickly replaced by the orthogonal
``hard wall'', the separat\-rix
plane~($ x, z $) in
Figure~\ref{fig12}c.
%
%%% Figure 12 (c) %%%
%
Thus the topological trigger drastically changes directions of
magnetic fluxes in an AR as illustrated by
Figure~14.
%%% Figure~\ref{fig14}. %%%
%
%%%%% Figure 14 %%%%%
%
%%%%%%%%%%%%%%%%%%%%%%%%%%%%%%%%%%%%%%%%%%%%%%%%%%%%%%%%%%%%%%%%%%%%%%%
\begin{figure} % [tbh]
%\hspace*{3mm}
\parbox[b]{96mm}{ % Left box %%%%%
    \epsfxsize=96mm %%%
    \epsfbox{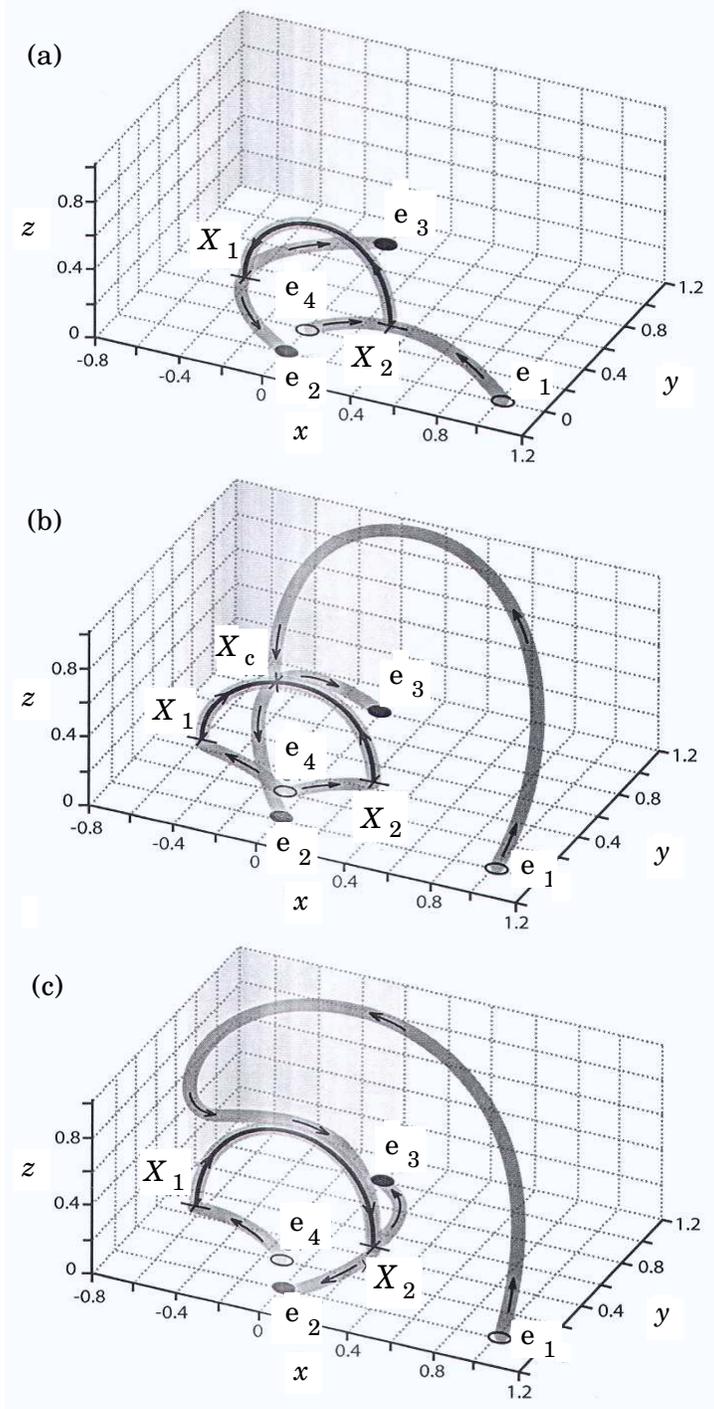}
%     \label{fig14}
    } %%%%%
    \hspace{3mm}
\parbox[b]{46mm}{ % Right box %
    \caption{Global changes of the magnetic field configuration
             related to the topological trigger effect in the
             model of an active region as illustrated by
             Figure~13. %%%
   (a) An initial state before the trigger; the zeroth
       point~$ X_{1} $ is the non-degenerate point
       with a vertical component of magnetic field $ B_{z}> 0 $.
   (b) After beginning of the trigger, the point~$ X_{1} $ is
       the non-degenerate point
       with $ B_{z}< 0 $; there are two zeroth points outside
       the plane~$ z = 0 $, one of them is shown as $ X_{c} $
       above the plane~$ z = 0 $.
   (c) The end of the topological trigger process.}
    \vspace{13mm}
    \noindent %%%
    }
  \label{fig14}
\end{figure}

Another specific sample is a charge arrangement along a straight line,
e.g., the axis~$ x $ in
Figure~\ref{fig1}.
%
%%% Figure 1 %%%
%
In this case, owing to the axial symmetry, the entire separator
consists of zeroth points.
Thus we have a zeroth line.
By using an inversion transformation (e.g.
Landau et al.~[38]):
%%% Landau et al.,~1984): %%%
% (32)
\begin{equation}
    {\bf r}^{\prime} = \frac{ R^{2} }{ r^{2} } \, {\bf r} \, ,
    \label{080103}
\end{equation}
where $ R $ is the inversion radius, it is easy to show that the
axial symmetry is not a necessary condition for the appearance of
zeroth line.
Moreover, the new zeroth line also represents a circle centered in the
plane~$ z = 0 $.
Thus the 3D zeroth lines of the magnetic field can exist if the
sunspots do not lie in a straight line.

Up to now, we have considered the travel of one charge while the
coordinates and magnitudes of the other three charges were fixed.
It is obvious, however, that all the foregoing remains in force in
the more general case of variation of the charge configuration.
It follows from the results that a slow evolution of the configuration
of field sources in the photosphere can lead to a rapid
rearrangement of the global topology in ARs in
the corona.
The phenomenon of topological trigger is necessary to
model the large eruptive flares.

%
%%% Section 4 %%%   %%%   %%%   %%%   %%%%%%%%%%%%%%%%%%%%%%%%%%%%%%%%%%
%
\section{Basic physics of reconnection in the corona}
   \label{sec:BPoH}

%
%%% Sub-section 4.1 %%%   %%%%%%%%%%%%%%%%%%%%%%%%%%%%%%%%%%%%%%%%%%%%%%
%
\subsection{Super-hot turbulent-current layers}
   \label{sub:Gf}

Coulomb collisions do not play any role in the
{\em super-hot turbulent-current layers\/} (SHTCL).
So the plasma inside the SHTCL has to be considered as
col\-li\-sion\-less
(see Somov~[19]).
%%% (see So\-mov,~2006). %%%
The concept of an anomalous resistivity,
which originates from wave-particle interactions, is then useful to
describe the fast conversion from field energy to particle energy.
Some of the general properties of such a
{\bf colli\-sion\-less reconnection\/}
can be examined in a frame of a self-consistent model which makes it
possible to estimate the main parameters of the SHTCL.
%
%%%%%% Figure 15 %%%%%%
%
\begin{figure}[htb]
\vspace{1mm}
\epsfysize=49mm
   \centerline{\epsfbox{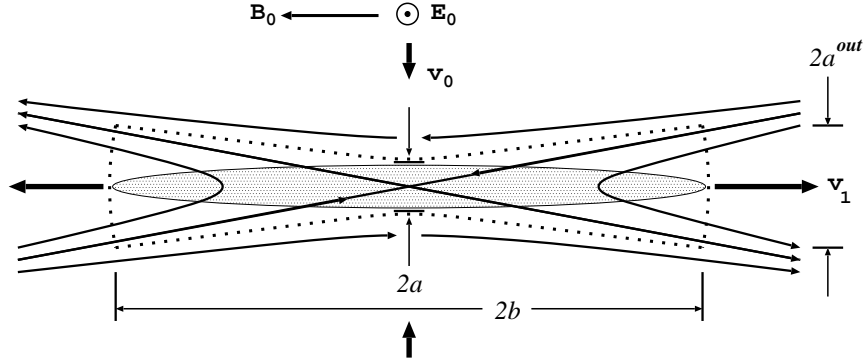}}
% \vspace{-1mm}
   \caption{A magnetically non-neutral reconnecting layer:
            the electric current distribution is schematically shown
            by the shadow, the dotted boundary indicates the field
            lines going through the current layer.
            $ 2a $ is the thickness of the current layer,
            $ 2b $ is its width.
            $ 2a^{out} $ is an effective cross-section for the outflows
            of energy and mass along the reconnected magnetic-field
            lines.
            }
   \label{fig15}
\end{figure}

As it reconnects, every magnetic-field line at first penetrates into
the current layer with velocity~$ {\bf v}_{0} $ and later on it moves
out with velocity~$ {\bf v}_{1} $ as
illustrated by
Figure~\ref{fig15}.
Basing on the mass, momentum and energy conservation laws,
we write the following relations:
% (33)
\begin{equation}
    n_0 v_0 \, b = n_s v_1 \, a^{out} ,
    \label{mass}
\end{equation}
% (34)
\begin{equation}
    2 n_0 k_{_{\rm B}} T_0
  + \frac{ B_0^{\, 2} }{ 8 \pi }
  = n_s k_{_{\rm B}} T
    \left( 1 + \frac{1}{\theta} \right) ,
    \label{momentum1}
\end{equation}
% (35)
\begin{equation}
    n_s k_{_{\rm B}} T \left( 1
  + \frac{1}{\theta} \right)
  = \frac{1}{2} \, M n_s v_1^{\, 2} + 2 n_0 k_{_{\rm B}} T_0 \, ,
    \label{momentum2}
\end{equation}
% (36)
\begin{equation}
    \chi_{\rm ef} \, {\cal E}_{mag}^{\, in}
  + {\cal E}_{\, th,e}^{\, in}
  = {\cal E}_{\, th,e}^{\, out}
  + {\cal C}_{\, \parallel}^{\, an} \, ,
    \label{energye}
\end{equation}
% (37)
\begin{equation}
    (1 - \chi_{\rm ef} ) \, {\cal E}_{mag}^{\, in}
  + {\cal E}_{\, th,i}^{\, in} = {\cal E}_{\, th,i}^{\, out}
  + {\cal K}_{i}^{\, out} \, .
    \label{energyi}
\end{equation}
Here $ n_0 $ and $ n_s $ are the plasma densities outside and inside
the layer.
$ T_0 $ is the temperature of inflowing plasma outside the layer,
$ T = T_e $ is an effective electron temperature inside it, the
ratio~$ \theta = T_e / T_i $,
$ T_i $ is an effective temperature of ions.
The velocity of electric drift of plasma in direction to the layer
% (38)
\begin{equation}
    v_0 = V_d = c \, \frac{ E_{0} }{ B_{0} } \, ,
    \label{Vd}
\end{equation}
and
% (39)
\begin{equation}
    v_1
  = V_{_{ \rm A,S }} = \frac{ B_0 }{ \sqrt{ 4 \pi M n_s } }
    \label{VAs}
\end{equation}
is the velocity of the plasma outflow.

The continuity Equation~(\ref{mass}) and the energy
Equations~(\ref{energye}) and~(\ref{energyi}) are of integral form
for a quarter of the layer assumed to be symmetrical and for
a unit length along the current.
The left-hand sides of (\ref{energye}) and~(\ref{energyi}) contain the
magnetic energy flux
% (40)
\begin{equation}
    {\cal E}_{mag}^{\, in}
  = \frac{ B_0^{\, 2} }{ 4 \pi } \, v_0 \, b \, ,
    \label{enthalpym}
\end{equation}
which creates heating of electrons and ions due to their interactions
with waves.
A relative fraction~$ \chi_{\rm ef} $ of the heating is consumed by
electrons, while the remaining fraction~$ ( 1 - \chi_{\rm ef} ) $
goes to the ions.

The electron and ion temperatures of the plasma inflowing to the
layer are the same.
Hence, the fluxes of the electron and ion thermal energies are
also the same:
% (41)
\begin{equation}
    {\cal E}_{\, th, e}^{\, in}
  = {\cal E}_{\, th, i}^{\, in}
  = \frac{5}{2} \, n_0 k_{_{\rm B}} T_0 \cdot v_0 b \, .
    \label{enthalpies}
\end{equation}
The factor 5/2 appears because, for the particles of kind~$ k $
(electrons, protons, other ions), the mean kinetic energy flux of
chaotic motion,~$ n_{k} m_{k} v_{k} \, w_{k} $,
where the heat function per unit mass or, more exactly, the
specific enthalpy is
\begin{displaymath}
    w_{k} = \varepsilon_{k} + \frac{ p_{k} }{ \rho_{k} }
  = \frac{5}{2} \, \frac{ k_{_{\rm B}} T_{k} }{ m_k } \, .
    \label{specific}
\end{displaymath}

Because of the difference between the temperatures of
electrons and ions in the out\-flowing plasma, the electron
and ion thermal energy outflows differ:
% (42)
\begin{equation}
    {\cal E}_{\, th,e}^{\, out} = \frac{5}{2} \,
    n_s k_{_{\rm B}} T \cdot v_1 a^{out} \, , \quad \,\,
    {\cal E}_{\, th,i}^{\, out} = \frac{5}{2} \,
    n_s k_{_{\rm B}} \frac{T}{ \theta } \cdot v_1 a^{out} \, .
\end{equation}
The ion kinetic energy flux from the layer
% (43)
\begin{equation}
    {\cal K}_{i}^{\, out}
  = \frac{1}{2} \, M n_s v_1^{\, 2} \cdot v_1 a^{out}
\end{equation}
is important in the energy balance~(\ref{energyi}).
As to the electron kinetic energy, it is negligible and disregarded
in (\ref{energye}).
However, electrons play the dominant role in the
heat conductive cooling of a SHTCL:
% (44)
\begin{equation}
    {\cal C}_{\, \parallel}^{\, an} = f_{_{\rm M}} ( \theta ) \,
    \frac{ n_s ( k_{_{ \rm B}} T )^{3/2} }{ M^{1/2} } \, a^{out} .
    \label{Manh}
\end{equation}
Here $ f_{_{\rm M}} ( \theta ) $
is the function which allows us to consider the field-aligned
{\em anomalous\/} thermal flux depending on the the ratio~$ \theta $.

Under the conditions derived from the {\it Yohkoh\/}
and {\it RHESSI\/} data, contributions to the energy balance are not
made either by the energy exchange between the electrons and the ions
due to collisions, the thermal flux across the magnetic field, and
the energy losses for radiation.
The field-aligned thermal flux becomes anomalous and plays
the dominant role in the cooling of electron component inside the
SHTCL.
All these properties are typical for col\-li\-sion\-less
`{\em super-hot\/}'
($ T_{e} \stackrel{>}{_\sim} 30 $~MK) plasma.

Under the same conditions, the
anomalous conductivity~$ \sigma_{\rm ef} $ in the Maxwell equation
for~$ {\rm curl} \, {\bf B} $
% (45)
\begin{equation}
    \frac{ c B_0 }{ 4 \pi a } = \sigma_{\rm ef} E_0 \, ,
    \label{cB0}
\end{equation}
as well as the relative fraction~$ \chi_{\rm ef} $ of the direct
heating consumed by electrons, are determined by the wave-particle
interaction inside the SHTCL and depend on a type of plasma
turbulence and its regime.

%
%%% Sub-section 4.2 %%%   %%%%%%%%%%%%%%%%%%%%%%%%%%%%%%%%%%%%%%%%%%%%%%
%
\subsection{Plasma turbulence inside the SHTCL} %%%%%%%%%%%%%%
   \label{sub:Psfa}

In Equation~(\ref{cB0}) it is convenient to replace the effective
conductivity~$ \sigma_{\rm ef} $ by effective
resistivity~$ \eta_{\, \rm ef} $:
% (46)
\begin{equation}
    \frac{ c B_0 }{ 4 \pi a }
  = \frac{ E_0 }{ \eta_{\, \rm ef} } \, .
    \label{1cB0}
\end{equation}

In general, the partial contributions to the resistivity
may be made simultaneously by several processes of electron
scattering by different sorts of waves, so that
% (47)
\begin{equation}
    \eta_{\, \rm ef} = \sum_{k} \eta_{\, k} \, .
    \label{etaef}
\end{equation}
The relative share of the electron heating~$ \chi_{\rm ef} $ is also
presented as a sum of the respective shares~$ \chi_{k} $ of the
feasible processes taken, of course, with the weight
factors~$ \eta_{\, k} / \eta_{\, \rm ef} $ which defines the relative
contribution from one or another process to the total heating of
electrons inside the SHTCL:
% (48)
\begin{equation}
    \chi_{\rm ef}
  = \sum_{k} \frac{ \eta_{\, k} }{  \eta_{\, \rm ef} } \,
    \chi_{k} \, .
    \label{sumchi}
\end{equation}

In usual practice
(e.g. So\-mov~[26]),
%%% (e.g., Somov,~1992), %%%
the sums~(\ref{etaef}) and~(\ref{sumchi}) consist of no more than two
terms, either of which corresponds to one of the turbulent types or
states.

%
%%% Sub-section 4.3 %%%   %%%%%%%%%%%%%%%%%%%%%%%%%%%%%%%%%%%%%%%%%%%%%%
%
\subsection{Marginal and saturation regimes}
   \label{sub:Pti}

When the electron current velocity~$ u = j / n_{s} $ exceeds a
critical value, the instabilities due to current flow of electrons
appear.
A rapid decrease in the plasma conductivity occurs and
anomalous resistivity arises
(Kadomtsev~[39],
Artsimo\-vich and Sag\-deev~[40]).
%%% (Kadom\-tsev,~1976; %%%
%%% Artsimo\-vich and Sagdeev,~1979). %%%
The condition needed for current instability in a RCL is that the
layer thickness is of the order of the ion gyro\-radius
(Sy\-ro\-vat\-skii~[17]).
%%% (Syro\-vatskii,~1981). %%%
Thus the turbulent current layers must be sufficiently thin.
According to
Sy\-ro\-vat\-skii~[41],
%%% Syro\-vatskii~(1972), %%%
the development of a {\em thin current layers\/} (TCL) in the solar
atmosphere leads to plasma turbulence and correspondingly to a fast
rate of field dissipation in flares with the heating of plasma to high
temperatures, the high-velocity plasma ejections, and the acceleration
of particles to high energies.

\begin{figure}[thb]
   \epsfysize=70mm
\vspace{1mm}
   \centerline{\epsfbox{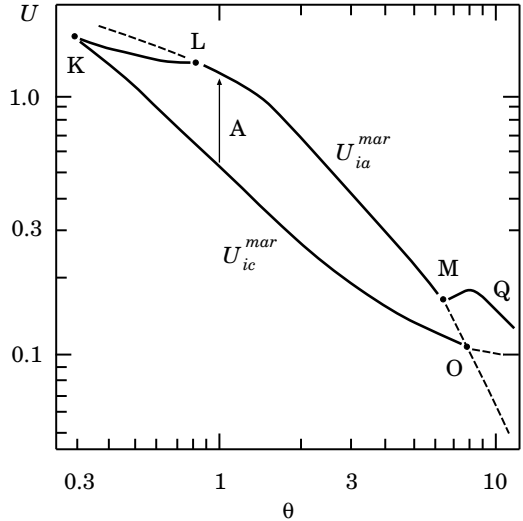}}
   \caption{The relative velocity of electrons, $ U = u / V_{Te} $ as
            a function of the ratio~$ \theta = T_{e} / T_{i} $ inside
            the SHTCL.
            The individual `arcs' of the curves correspond to
            four different regimes of turbulence.
            The dashed continuations of the `arcs' are the critical
            velocities of excitation of corresponding waves.
            The arrow~A shows that, in equilibrium plasma with
            $ \theta = 1 $, the ion-cyclotron waves (the curved
            section~KO) are excited earlier than the ion-acoustic
            waves (the arc~LO).
            }
   \label{fig16}
\end{figure}

Let us consider the turbulence which is due only to the
{\em ion-cyclotron\/} ($ ic $) and {\em ion-acoustic\/} ($ ia $)
instabilities.
The sums~(\ref{etaef}) and~(\ref{sumchi}) consist of no more than
two terms; either of which corresponds to one of the said turbulence
types.
In reality, the ion-cyclotron waves prove to be excited earlier than
the ion-acoustic waves at all values of the temperature
ratio~$ \theta \stackrel{ < }{_\sim} 8 $
(Figure~\ref{fig16}).
%
%%% Figure 16 %%%
%

%
%%%%%% Figure 16 %%%%%%   %%%%%%%%%%%%%%%%%%%%%%%%%%%%%%%%%%%%%%%%%%%%%%
%

In the case of the {\em marginal\/} regime of turbulence, the velocity
of electrons
% (49)
\begin{equation}
    u
  = \frac{ E_0 }{ e n_s \, \eta_{\, \rm ef} }
    \label{uelectrons}
\end{equation}
coincides with the critical velocity~$ u_k $ of the $ k $-type wave
excitation.
Hence
% (50)
\begin{equation}
    U_{k } \, ( \theta ) = U_{k }^{\, mar} \, ( \theta )
  = \frac{ u_k }{ V_{_{\rm Te}} } \,\, .
    \label{Uktheta}
\end{equation}
For example, the ion-cyclotron instability becomes enhanced when the
velocity~$ u $ is not lower than the critical
value~$ u_{ic} $ of the ion-cyclotron waves.
In the marginal regime
% (51)
\begin{equation}
    U_{ic }^{\, mar} \, ( \theta )
  = \frac{ u_{ic} }{ V_{_{\rm Te}} } \,\, .
    \label{Ukic}
\end{equation}
This function is shown as the curved section~KO in
Figure~\ref{fig16}.
%
%%% Figure 16 %%%
%

As long as the ion-cyclotron waves are not saturated, the
velocity~$ u $ remains approximately equal to~$ u_{\, ic} $
and thus it is possible to calculate the
resistivity~$ \eta_{\, \rm ef} $ from
Equation~(\ref{uelectrons}).
Therefore, if the velocity~$ u $ in the SHTCL
does not exceed the ion-acoustic wave excitation
threshold~$ u_{\, ia}$,
only the ion-cyclotron waves will contribute to anomalous
resistivity~$ \eta_{\, \rm ef} $
and the factor~$ \chi_{\, \rm ef} $.

In the marginal regime, the wave-particle interaction
is {\em quasi\-linear\/}.
However, in the case of sufficiently strong electric
field~$ E_{0} $, the {\em nonlinear\/} interactions
become important, thereby giving rise to another state of
turbulence.
In this regime, Ohm's law can no longer define the resistivity, but
determines the the electron current velocity.
Regarding the resistivity, this is inferred from the turbulence
saturation level.
In the {\em saturated\/} regime,
$  U_{\, k } \, ( \theta ) $ must be
replaced by certain functions~$ U_{ic}^{\, sat} ( \theta )  $,
shown by the curved section~KL in
Figure~\ref{fig16},
%
%%% Figure 16 %%%
%
and $ U_{ia}^{\, sat} ( \theta ) $, shown by the arcs~MQ for the
ion-cyclotron and ion-acoustic turbulence, respectively.

Calculations show that the energy release power and the reconnection
rate, which are necessary for solar flares to be
accounted for, can be obtained in the marginal regime of
ion-acoustic turbulence (see
So\-mov~[26,19]).
%%% Somov,~1992, 2006). %%%

%%%%%%%%%%%%%%%%%%%%%%%%%%%%%%%%%%%%%%%%%%%%%%%%%%%%%%%%%%%%%%%%%%%%%%%%
%
%%% Section 5 %%%   %%%  %%%  %%%   %%%%%%%%%%%%%%%%%%%%%%%%%%%%%%%%%%%%
%
\section{The collapsing magnetic trap effects} %%%
    \label{sec:3mech}

%
%%% Sub-section 5.1 %%%%%%%%%%%%%%%%%%%%%%%%%%%%%%%%%%%%%%%%%%%%%%%%%%%%
%
\subsection{Fast and slow reconnection}
   \label{sub:FaSr}

\begin{figure} % [hb]
    \epsfxsize=130mm
    \centerline{\epsfbox{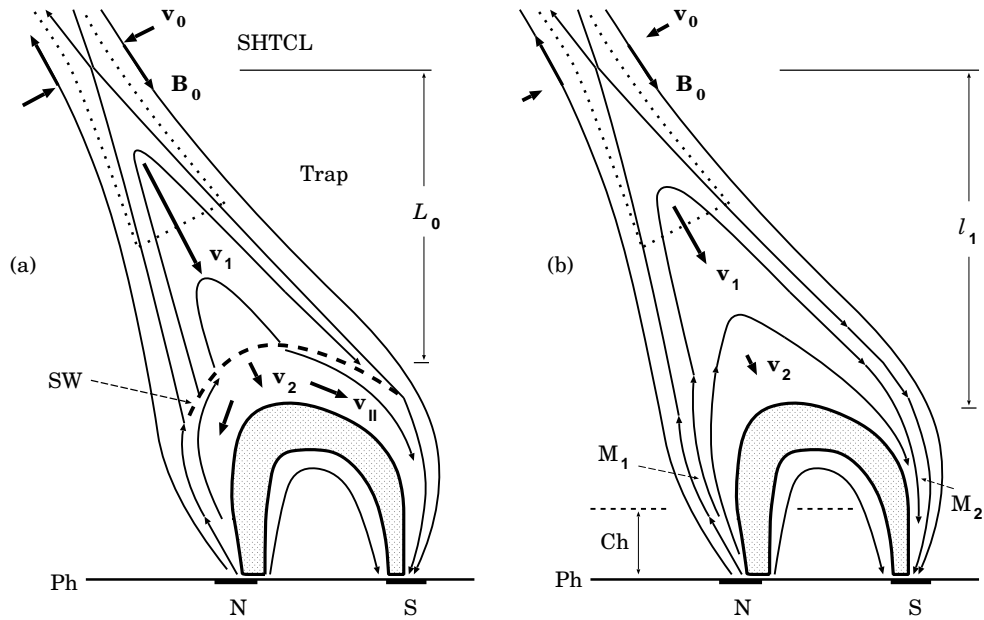}}
    \caption{Plasma flows related to a super-hot turbulent-current
        layer (SHTCL): the inflows with a relatively low
        velocity~$ {\bf v}_0 $, the downward outflow with a
        super-Alfv\'{e}n velocity~$ {\bf v_1} $.
  (a) SW is the shock wave above the magnetic obstacle.\,
        $ {\bf v}_2 $ is the post-shock
        velocity, $ {\bf v}_{\parallel} $ is the velocity of
        spreading of the compressed plasma along the field lines
        toward the feet of the loop.
  (b) The supra-arcade down\-flow and collapsing trap without a
        shock. M$_{1}$ and M$_{2}$ are the mirroring points where the
        field becomes sufficiently strong to reflect fast particles
        above the chromosphere~(Ch).}
    \label{fig17}
\end{figure}

Collapsing magnetic traps are formed by the process of
collision\-less reconnection in the solar atmosphere
(So\-mov and Ko\-sugi~[42]).
%%% (Somov and Kosugi,~1997). %%%
Figure~\ref{fig17}
%
%%% Figure 15 %%%
%
illustrates two possibilities.
Fast (Figure~\ref{fig17}a) and
slow (Figure~\ref{fig17}b) modes of reconnection are sketchy
shown in the corona above the magnetic obstacle, the region of a strong
magnetic field, which is observed in SXRs as a flare loop (shaded).

%
%%%%%% Figure 17 %%%%%%   %%%%%%%%%%%%%%%%%%%%%%%%%%%%%%%%%%%%%%%%%%%%%%
%

In the first case, let us assume that both feet of a reconnected field
loop path through the shock front (SW in
Figure~\ref{fig17}a)
%
%%% Figure 17 %%%
%
ahead the obstacle.
Depending on the velocity and pitch-angle, some of the particles
pre\-accelerated by the SHTCL may penetrate through the magnetic-field
jump related to the shock or may be reflected.
For the particles reflected by the shock, the magnetic loop represents a
trap whose length~$ L (t) $, the distance between two mirroring points
at the shock front, measured along a magnetic-field line, decreases from
its initial value
$ L (0) \approx 2 L_{0} $ to zero (the top of the loop goes through the
shock front) with the velocity~$ \approx 2 v_{1} $.
Therefore, the lifetime of each collapsing trap
$ t_{1} \approx L_{0} / v_{1} $.

In the case of slow reconnection, there is no a shock wave, and the trap
length~$ L (t) $ is the distance between two mirroring points
(M$ _{1} $ and M$ _{2} $ in
Figure~\ref{fig17}b),
%
%%% Figure 15 %%%
%
measured along a reconnected magnetic-field line.
In both cases, the electrons and ions are captured in a trap whose
length decreases.
So the particles gain energy from the increase in parallel momentum.

\begin{figure}[hb]
    \epsfxsize=152mm
    \centerline{\epsfbox{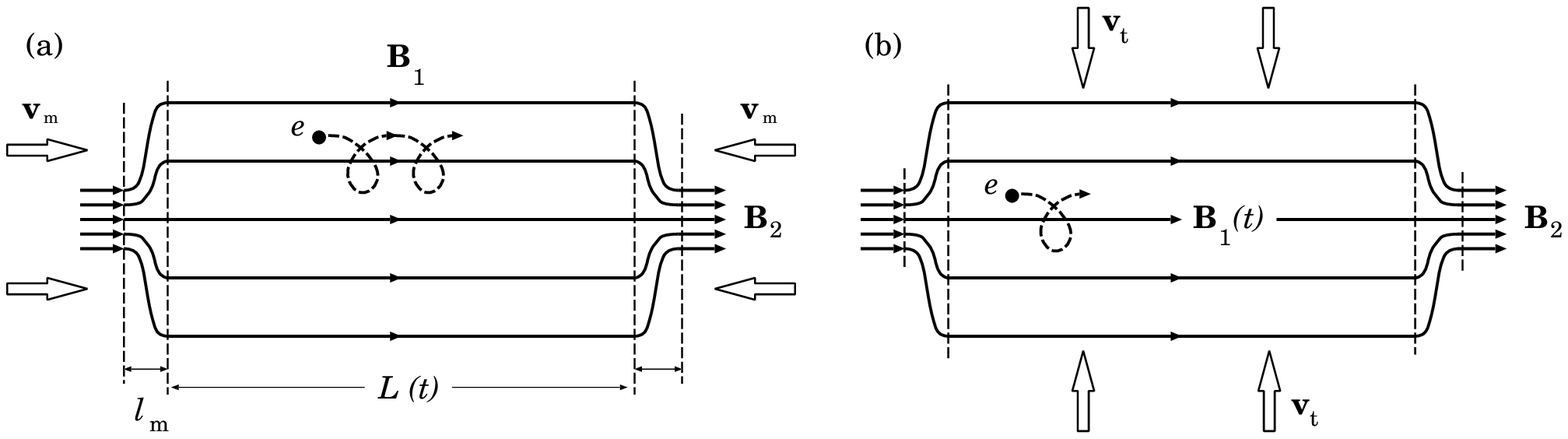}}
    \caption{Two main effects in a collapsing trap.
    (a) Magnetic mirrors move toward each other with
            velocity~$ {\bf v}_{m} $.
    (b) Compression of the trap with velocity~$ {\bf v}_{t} $.}
    \label{fig18}
\end{figure}

Thus, in the first approximation, we neglect collisions of particles
ahead of the shock wave
(Figure~\ref{fig17}a)
%
%%% Figure 17 %%%
%
or in the trap without a shock
(Figure~\ref{fig17}b).
%
%%% Figure 17 %%%
%
In both cases, the particle acceleration can be demonstrated in a
simple model -- a long trap with short mirrors
(Figure~\ref{fig18}).
%
%%%%% Figure 18 %%%
%
%
%%%%%% Figure 18 %%%
%
The decreasing length~$ L (t) $ of the trap is much larger than the
length~$ l_{m} $ of the mirrors; the magnetic
field~$ {\bf B} = {\bf B}_{1} $ is uniform inside the trap but grows
from $ {\bf B}_{1} $ to $ {\bf B}_{2} $ in the mirrors.
The quantity~$ B_{2} / B_{1} $ is called the mirror ratio; the larger
this ratio, the higher the particle confinement in the trap.
The validity conditions for the model are discussed by
%%%
So\-mov and Bo\-ga\-chev~[43].
%%% So\-mov and Bo\-ga\-chev~(2003). %%%

%
%%% Sub-section 5.2 %%%   %%%%%%%%%%%%%%%%%%%%%%%%%%%%%%%%%%%%%%%%%%%%%%
%
\subsection{The first-order Fermi-type acceleration}
   \label{sub:Fermi}

We consider the traps for those the length scale and timescale are
both much larger than the gyro\-radius and gyro\-period of an
accelerated particle.
Due to strong separation of length and timescales, the magnetic field
inside the trap can be considered as uniform and constant
(for more detail see
So\-mov and Bo\-ga\-chev~[43]).
%%% So\-mov and Bo\-ga\-chev,~2003). %%%
If so, then the longitudinal momentum of a particle increases with
a decreasing length~$ L (t) $, in the adiabatic approximation, as
% (52)
\begin{equation}
    \ppr(l) = \frac{ \popr }{ l } \, .
    \label{ppr1}
\end{equation}
Here $ l = L (t) / L (0) $ is the dimensionless length of the trap.
The transverse momentum is constant inside the trap,
% (53)
\begin{equation}
    \ppp = \popp \, ,
    \label{ppp}
\end{equation}
because the first adiabatic invariant is conserved:
% (54)
\begin{equation}
    \frac{ \ppp^{\,2} }{ B } = {\rm const} \, .
    \label{ai}
\end{equation}
Thus the kinetic energy of the particle increases as
% (55)
\begin{equation}
    { K} (l)
  = \frac{ \ppr^{\, 2} + \ppp^{\, 2}}{2\,m}
  = \frac{1}{2\,m}
    \left(
    \frac{\popr^{\, 2}}{l^{\, 2}}
  + \popp^{\, 2}
    \right) .
    \label{Ek1}
\end{equation}

The time of particle escape from the trap, $ l = l_{es} $, depends on
the initial pitch-angle~$ \theta_{0} $ of the particle and is
determined by the condition
% (56)
\begin{equation}
    {\rm tg} \, \theta_{0}
  = \frac{ \popp }{ \popr }\, \le
    \frac{ 1 }{ R \, l_{es} } \, ,
    \label{llim}
\end{equation}
where
% (57)
\begin{equation}
    R = \left( \frac{ B_{\, 2} }{ B_{\, 1} } - 1 \right)^{1/2} \, .
    \label{R-first}                                    %%% R-first %%%
\end{equation}
The kinetic energy of the particle at the time of its escape is
% (58)
\begin{equation}
    { K}_{es} =
    \frac{ \popp^{\, 2} }{ 2\,m } \, \left( R^{2} + 1 \right)
  = \frac{ \popp^{\, 2} }{ 2\,m} \,
    \frac{ B_{\, 2} }{ B_{\, 1} } \, .
    \label{Ekc}
\end{equation}
One can try to obtain the same canonical result by using more
complicated approaches.
For example,
Giuliani et al.~[44]
%%% Giuli\-ani et al.~(2005) %%%
numerically solved the drift equations of motion.
However it is worthwhile to explore first the simple analytical
approach to investigate the particle
energi\-zation processes in collapsing traps in more detail
before starting to use more sophisticated methods and simulations.

%
%%% Sub-section 5.3 %%%   %%%%%%%%%%%%%%%%%%%%%%%%%%%%%%%%%%%%%%%%%%%%%%
%
\subsection{The betatron acceleration in a collapsing trap}
   \label{sub:beta}

If the thickness of the trap decreases with its decreasing
length, then the strength of the field~$ {\bf B}_{1} $ inside the
trap increases as a function of
$ l $, say $ B_{\, 1} (l) $.
In this case, according to (\ref{ai}), the transverse momentum
increases simultaneously with the longitudinal momentum~(\ref{ppr1}):
% (59)
\begin{equation}
    \ppp(l)
  = \popp \, \left(
    \frac{ B_{\, 1} (l) }{ B_{\, 1} } \right)^{1/2}\, .
    \label{ppl}
\end{equation}
Here $ B_{\, 1} = B_{\, 1} (1) $ is the initial (at $ l = 1 $) value
of the field inside the trap.
The kinetic energy of a particle
% (60)
\begin{equation}
    { K} (l)
  = \frac{1}{2\,m}
    \left(
    \frac{\popr^{\, 2}}{l^{\, 2}}
  + \popp^{\, 2} \,
    \frac{ B_{\, 1} (l) }{ B_{\, 1} }
    \right)
\end{equation}
increases faster than that in the absence of trap contraction, see
(\ref{Ek1}).
Therefore it is natural to assume that the acceleration efficiency in
a collapsing trap also increases.

%
%%%%%% Figure 19 %%%%%%   %%%%%%%%%%%%%%%%%%%%%%%%%%%%%%%%%%%%%%%%%%%%%%
%
\begin{figure}[h]
    \epsfxsize=148mm
    \centerline{\epsfbox{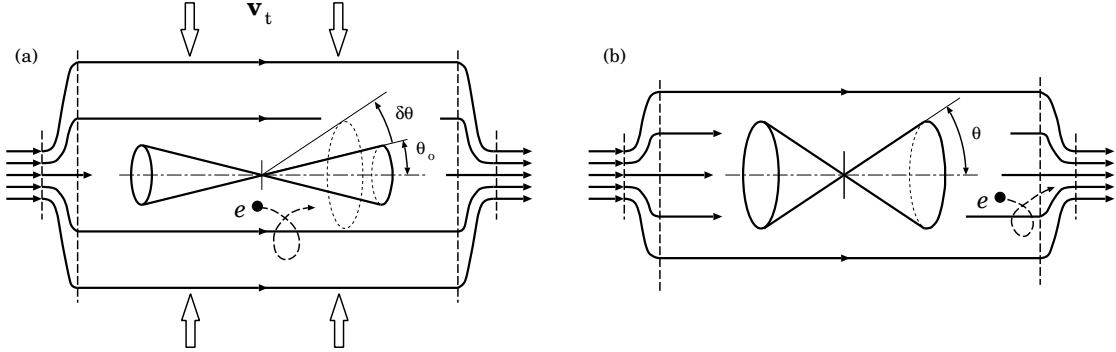}}
    \caption{The betatron effect in a collapsing magnetic trap.
    As the trap is compressed with
    velocity~$ {\bf v}_{t} $, the loss cone becomes larger.
    A particle escapes from the trap earlier with an
    additional energy due to betatron acceleration.}
    \label{fig19}
\end{figure}

However, as the trap is compressed, the loss cone becomes larger
(Figure~\ref{fig19}),
%
%%% Figure 19 %%%
%
% (61)
\begin{equation}
    \theta_{es} (l)
  = {\rm arcsin} \,
    \left( \frac{ B_{\, 1} (l) }{ B_{\, 2} } \right)^{1/2} \, .
\end{equation}
Consequently, the particle escapes from the trap earlier.

On the other hand, the momentum of the particle at the time of its
escape satisfies the condition
% (62)
\begin{equation}
    {\ppr} (l) = {R} (l) \, {\ppp} (l) \, ,
\end{equation}
where
% (63)
\begin{equation}
    R(l) =\left(
    \frac{ B_{\, 2} }{ B_{\, 1} (l) } - 1 \right)^{1/2} \, .
    \label{R-lasr}                                      %%% R-last %%%
\end{equation}
Hence, using~(\ref{ppl}), we determine the energy of the particle at
the time of its escape from the trap
% (64)
\begin{equation}
    { K}_{es}
  = \frac{ \ppp (l)^{\, 2} }{ 2 \, m }
    \left( R (l)^{\, 2} + 1 \right)
  = \frac{ \popp^{\,2} }{2\,m} \,
    \frac{ B_{\, 1} (l) }{ B_{\, 1} } \,
    \frac{ B_{\, 2} }{ B_{\, 1} (l) } \,
  = \frac{ \popp^{\, 2} }{2\, m} \,
    \frac{ B_{\, 2} }{ B_{\, 1} } \, .
    \label{Ekk}
\end{equation}
The kinetic energy~(\ref{Ekk}), that the particle gains in a
collapsing trap with compression, is equal to the energy~(\ref{Ekc})
in a collapsing trap without compression, i.e. without the betatron
effect.

Thus the compression of a collapsing trap (as well as its expansion or
the transverse oscillations) does not affect the final energy that the
particle acquires during its acceleration.
The faster gain in energy is {\em exactly\/} offset by the earlier
escape of the particle from the trap
(So\-mov and Bo\-ga\-chev~[43]).
%%% (So\-mov and Bo\-ga\-chev,~2003). %%%
The acceleration efficiency, which is defined as the ratio of the
final
$ ( l = l_{ls} ) $ and initial $ ( l = 1 ) $ energies, i.e.
% (65)
\begin{equation}
    \frac{ { K}_{es} }{ { K} (1) }
  = \frac{ \popp^{\, 2} }{ \popp^{\, 2} + \popr^{\, 2} } \,
    \frac{ B_{\, 2} }{ B_{\, 1} }
  = \left(
    \frac{ \popp }{ p_{\, 0} }
    \right)^{\! 2} \,
    \frac{ B_{\, 2} }{ B_{\, 1} } \, ,
    \label{eff}
\end{equation}
depends only on the initial mirror
ratio~$ B_{\, 2} / B_{\, 1} $ and the initial particle momentum or, to
be more precise, on the ratio~$ \popp / p_{\, 0} $.
%%% Therefore, an initial anisotropy of accelerated particles can be
%%% essential ... .
The acceleration efficiency~(\ref{eff}) does not depend on the
compression of collapsing trap and the pattern of decrease in the
trap length either.

It is important that
the acceleration time in a collapsing magnetic trap with compression
can be much shorter than that in a collapsing trap without compression.
For example, if the cross-section area~$ S (l) $ of the trap decreases
proportionally to its length~$ l $:
% (66)
\begin{equation}
    S (l) = S (1) \, l \, ,
\end{equation}
then the magnetic field inside the trap
% (67)
\begin{equation}
    B_{\, 1} (l) = B_{\, 1} (1) \,/\, l \, ,
\end{equation}
and the effective parameter
% (68)
\begin{equation}
    R (l) =
    \left( R^{2}
  - \frac{ 1 - l }{ l } \right)^{1/2} \, ,
\end{equation}
where $ R $ is define by formula~(\ref{R-first}).
%%% R-first (7.15) ??? %%%
At the critical length
% (69)
\begin{equation}
    l_{cr} =
    \frac{ 1 }{ 1 + R^{2} } \, ,
\end{equation}
the magnetic field inside the trap becomes equal the field in the
mirrors, and the magnetic reflection ceases to work.
If, for certainty, $ B_{\, 2} / B_{\, 1} = 4 $, then
$ l_{cr} = 1/4 $.
So contraction of the collapsing trap does not change the energy of
the escaping particles but this energy is reached at an earlier stage
of the magnetic collapse when the trap length is finite.
In this sense, the betatron effect increases the actual
efficiency of the main process -- the particle acceleration on the
converging magnetic mirrors.

%
%%% Sub-section 5.4 %%%   %%%%%%%%%%%%%%%%%%%%%%%%%%%%%%%%%%%%%%%%%%%%%%
%
\subsection{The betatron acceleration in a shockless trap}
   \label{sub:less}

If we ignore the betatron effect in a shock\-less collapsing trap,
show in
Figure~\ref{fig17}b,
%
%%% Figure 15 %%%
%
then the longitudinal momentum of a particle is defined by the
formula
(instead of (\ref{ppr1}))
% (70)
\begin{equation}
    p_{\, \parallel} (t)
    \approx
    p_{\, \parallel} (0) \,
    \frac{ (l_1 + l_2) }{ l_2 + (l_1 - v_1 t) }
    \Rightarrow
    p_{\, \parallel} (0) \,
    \frac{ (l_1 + l_2 ) }{ l_2 }
    \, , \quad {\rm when} \quad
    t \rightarrow t_1 \, .
    \label{1ppar-2}
\end{equation}
The particle acceleration on the magnetic mirrors stops at the
time~$ t_{1} = l_{1} / v_{1} $ at a finite longitudinal momentum that
corresponds to a residual length ($ l_{2} $ in
Figure~\ref{fig17}b)
%
%%% Figure 15 %%%
%
of the trap.

Given the betatron acceleration due to compression of the trap, the
particle acquires the same energy~(\ref{Ekc}) by this time or earlier
if the residual length of the trap is comparable to a critical
length~$ l_{cr} $ determined by a compression law
(Somov and Bogachev~[43]).
%%% (So\-mov and Bo\-ga\-chev,~2003). %%%
%%% \cite[see][]{Somov03}. %%%
Thus the acceleration in shock\-less collapsing traps with a residual
length becomes more plausible.
The possible observational manifestations of such traps in the X-ray
and optical radiation are discussed by
Somov and Bogachev~[43].
%%% So\-mov and Bo\-ga\-chev (2003). %%%
The most sensitive tool to study behavior of the electron
acceleration in the collapsing trap is radio radiation.
We assume that wave-particle interactions are important and that
two kinds of interactions should be considered in the collapsing trap
model.

The first one is resonant scattering of the trapped electrons,
including the loss-cone instabilities and related kinetic processes
(e.g.
Benz~[45], Chapter~8).
%%% Benz~(2002), Chapter~8). %%%
Resonant scattering is most likely to enhance the rate of
precipitation of the electrons with energy higher that hundred keV,
generating microwave bursts.
The lose-cone instabilities of trapped mildly-relativistic electrons
(with account taken of the fact that there exist many collapsing field
lines at the same time, each line with its proper time-dependent loss
cone) would provide excitation of waves with a very wide continuum
spectrum.
In a flare with a slowly-moving upward coronal HXR source,
an ensemble of the collapsing field lines with accelerated
electrons would presumably be observed as a slowly moving type IV burst
with a very high brightness temperatures and with a possibly significant
time delay relative to the chromospheric foot\-point emission.

The second kind of wave-particle interactions in the collapsing
trap-plus-precipitation model is the streaming instabilities
(including the current instabilities related to a return current)
associated with the precipitating electrons.

%
%%% Sub-section 5.5 %%%   %%%%%%%%%%%%%%%%%%%%%%%%%%%%%%%%%%%%%%%%%%%%%%
%
\subsection{Some observational results}
    \label{sec:DaC+}

In order to interpret the temporal and spectral evolution and spatial
distribution of HXRs in flares, a two-step acceleration was
proposed by
So\-mov and Kosugi~[42]
%%% So\-mov and Ko\-su\-gi~(1997) %%%
with the second step of acceleration via the collapsing magnetic-field
lines.
The {\em Yohkoh\/} HXT observations of the Bastille-day flare
(Masuda et al.~[46])
%%% (Ma\-su\-da et al.,~2001) %%%
clearly show that, with increasing energy, the HXR emitting region
gradually changes from a {\em large diffuse source\/}, which is
located presumably above the ridge of soft X-ray arcade, to a
two-ribbon structure at the loop foot\-points.
This result suggests that electrons are in fact accelerated in the
large system of the coronal loops, not merely in a particular one.
This seems to be consistent with the {\em RHESSI\/} observations of
large coronal HXR sources; see, for example, the X4.8 flare of 2002
July~23 (see Figure~3 in
Lin et al.~[47]).
%%% Lin et al.,~2003). %%%

Efficient trapping and continuous acceleration also produce the large
flux and time lags of microwaves that are likely emitted
by electrons with higher energies, several hundred keV
(Kosugi et al.~[48]).
%%% (Ko\-su\-gi et al.,~1988). %%%
The lose-cone instabilities
(Benz~[45])
%%% (Benz,~2002) %%%
of trapped mildly-relativistic electrons in the system of many
collapsing field lines (each line with its proper time-dependent lose
cone) presumably can provide excitation of radio-wave with a very wide
continuum spectrum.

Qiu et al.~[49]
%%% Qiu et al.~(2004) %%%
presented a comprehensive study of the X5.6 flare on 2001 April~6.
Evolution of HXRs and microwaves during the gradual phase in this
flare exhibits a separation motion between two foot\-points,
which reflects the progressive reconnection.
The gradual HXRs have a harder and hardening spectrum compared with
the impulsive component.
The gradual component is also a microwave-rich event lagging
the HXRs by tens of seconds.
The authors propose that the collapsing-trap effect is a viable
mechanism that continuously accelerates electrons in
a low-density trap before they precipitate into the foot\-points.

Imaging radio observations (e.g.
Li and Gan~[50])
%%% Li and Gan,~2005) %%%
should provide another way to investigate properties of collapsing
magnetic traps.
It is not simple, however, to understand the observed phenomena
relative to the results foreseen by theory.
With the incessant progress of magnetic reconnection, the loop system
newly formed after reconnection will grow up, while every specific
loop will shrink.
Just because of such a global growth of flare loops, it is rather
difficult to observe the downward motion of newly formed loops.
The observations of radio loops by
No\-beyama Radio\-helio\-graph (NoRH)
are not sufficient to resolve specific loops.
What is observed is the whole region, i.e., the entire loop or the
loop top above it.
Anyway, combined microwave and HXR imaging observations are essential
in the future.

%
%%% Section 6 %%%   %%%  %%%  %%%   %%%%%%%%%%%%%%%%%%%%%%%%%%%%%%%%%%%%
%
\section{Open issues of reconnection in flares} %%%%%%%%%%%%%%%%%%%%%%%%
   \label{sec:OIRF}

\begin{figure} [hb]
   \epsfysize=60mm
\vspace{1mm}
   \centerline{\epsfbox{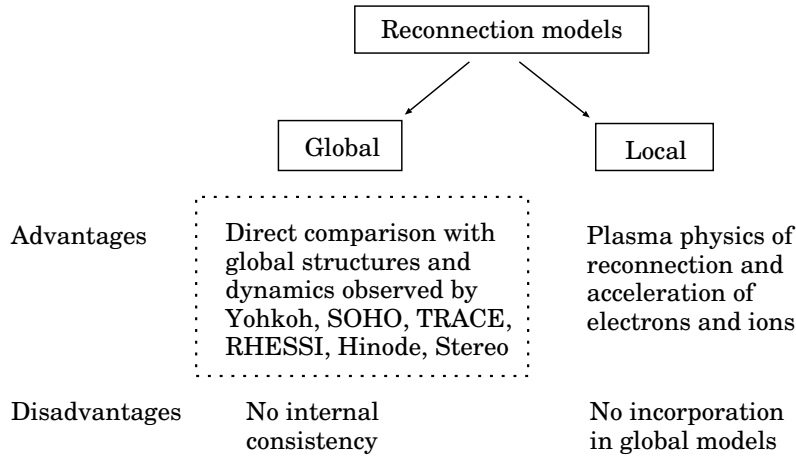}}
\vspace{1.3mm}
   \caption{Models of magnetic reconnection in the solar atmosphere.}
   \label{fig20}
\end{figure}

The existing models of reconnection in the solar corona can be
classified in two wide groups: global and local ones
(Figure~\ref{fig20}).
%
%%% Figure 20 %%%
%
%
%%%%%% Figure 20 %%%%%%   %%%%%%%%%%%%%%%%%%%%%%%%%%%%%%%%%%%%%%%%%%%%%%
%
The global models are used to describe actual ARs or even complexes
of activity in different approximations and with different accuracies
(So\-mov~[51],
Gorbachev and So\-mov~[7],
Bagal\'a et al.~[52],
Antiochos~[31],
Aschwanden et al.~[53],
Morita et al.~[54],
Longcope et al.~[55],
Barnes~[34],
Longcope and Beveridge~[56],
Ugarte-Urra et al.~[35]).
%%% (Somov,~1985; %%%
%%% Gorbachev and Somov,~1989; %%%
%%% Baga\-l\'a et al.,~1995; %%%
%%% Antio\-chos,~1998; %%%
%%% Aschwan\-den et al., 1999; %%%
%%% Mori\-ta et al.,~2001; %%%
%%% Longcope et al.,~2005; %%%
%%% Barnes,~2007; %%%
%%% Longcope and Beveridge,~2007; %%%
%%% Ugarte-Urra et al.,~2007). %%%
We make no attempt to review all the models but just remark that
the main advantage of the global models is direct comparison between
the results of computation and the observed large-scale patterns.
For example, the `{\em rainbow reconnection\/}' model
(So\-mov~[57])
%%% (Somov,~1986) %%%
is used to reproduce the main features of magnetic fields in the
corona, related to the large-scale {\em vortex\/} flows in the
photosphere.

The advantage of the local models is that they take kinetic effects
into account and allow us to develop the basic physics of reconnection
in flares.
In general, many analytical, numerical, and combined models of
reconnection exist in different approximations and with different
levels of self-consistency
(see Biskamp~[15],
Priest and For\-bes~[58]).
%%% (see Biskamp,~1997; %%%
%%% Priest and Forbes,~2000). %%%
It becomes more and more obvious that {\em collision\-less\/}
reconnection in a rarefied plasma is a key process in flares.
This process was introduced by
Syrovatskii~[59]
%%% Syro\-vatskii~(1966) %%%
as a {\em dynamic dissipation\/} of field in a current layer and leads
to fast conversion from field energy to particle energy, as well as a
topological change of the magnetic field
(e.g.
Horiuchi et al.~[60],
Yamada et al.~[16]).
%%% Horiuchi et al.,~2001; %%%
%%% Yamada et al.,~2007). %%%

General properties and parameters of the collision\-less reconnection
can be examined in a frame of models based on the mass,
momentum, and energy conservation laws
(Section~\ref{sec:BPoH}).
%
%%% Section 4 %%%
%
A particular feature of the models is that electrons and ions are
heated by wave-particle interactions in a different way.
The magnetic-field-aligned thermal flux becomes anomalous and plays
the role in the cooling of electrons in a
{\it super-hot turbulent-current layer\/}
(SHTCL).
These properties seem to be typical for the conditions derived
from the observations by {\em Yoh\-koh\/} and {\em RHESSI\/}
(e.g. Joshi et al.~[61],
Liu et al.~[62]),
%%% (e.g., Joshi et al.,~2007; %%%
%%% Liu et al.,~2008), %%%
the {\em SOXS\/} mission
(Jain et al.~[63]).
%%% (Jain et al.,~2006). %%%
Unfortunately, the local models are not incorporated in the global
consideration of reconnection in the corona.
Only a few first steps have been made in this direction (see
Somov~[19]).
%%% Somov,~2006). %%%

Spacecraft observations of collision\-less reconnection in the
magneto\-tail and magnetopause as well as recent simulations have
shown the existence of {\em thin\/} current layers (TCLs), with
scale lengths of the order a few electron skin depth.
In the electron MHD model of such TCLs, reconnection is facilitated
by electron inertia which breaks the frozen-in condition
(Jain and Sharma~[64]).
%%% (Jain and Sharma,~2007). %%%
The simulations demonstrate the instability of the whistler-like
mode which presumably plays the role of the ion-cyclotron instability
in the SHTCL model.
Stark et al.~[65]
%%% Stark et al.~(2007) %%%
combine the TCL models with a global model of magnetic field in the
magnetosphere.

As for particle acceleration in flares, it goes in two steps.
During the first one, the ions and electrons are accelerated by
the DC electric field in a super-hot turbulent-current layer
(Litvinenko and Somov~[66,67]).
%%% (Litvinenko and Somov,~1993, 1995). %%%
During the second step, the energy of particles trapped in collapsing
traps additionally increases by the first-order Fermi mechanism
(Somov and Kosugi~[42])
%%% (Somov and Kosugi,~1997) %%%
and the betatron mechanism
(Somov and Bogachev~[43],
Karlicky and Kosugi~[68]).
%%% (Somov and Bogachev,~2003; %%%
%%% Karlicky and Kosugi,~2004). %%%
In a collisionless approximation,
Bogachev and Somov~[69]
%%% Bogachev and Somov (2007) %%%
show that the collapsing trap model predicts two types of coronal
HXR sources: thermal and non-thermal.
Thermal sources are formed in traps dominated by the betatron
mechanism.
Non-thermal sources with power-law spectra appear when electrons
are accelerated by the Fermi mechanism.
These results are of interest in interpreting the {\em RHESSI\/}
observations.

Future models of solar flares should join global and local properties
of reconnection under coronal conditions.
For example, chains of plasma instabilities, including kinetic
instabilities, can be important for our understanding the types
and regimes of plasma turbulence inside the collision\-less
current layers with a longitudinal magnetic field.
In particular it is necessary to evaluate better the anomalous
resistivity and selective heating of particles in such a SHTCL.
Heat conduction is also anomalous in the super-hot plasma.
Self-consistent solutions of the reconnection problem will allow us to
explain the energy release in flares, including the open question of
the mechanism or combination of mechanisms which explains the observed
acceleration of electrons and ions to high energy.

To understand the 3D structure of reconnection in flares is
one of the most urgent problems.
Actual flares are 3D dynamic phenomenon of electromagnetic origin in
a highly-conducting plasma with a strong magnetic field.
{\em RHESSI\/} and {\em Hinode\/} observations offer us the
means to check whether phenomena predicted by topological models
(such as the topological trigger) do occur [70].
However some puzzling discrepancies may also exist, and further
development of realistic 3D models is required.

\vspace{1mm}

\section*{Acknowledgements}

The author thanks a reviewer for valuable
comments improving the text.\footnote{With minor differences in
Word/LaTeX setting of
formulae and references, this paper is published in:
{\em Asian Journal of Physics\/} {\bf 47},
421-454, 2008.}
This work was supported by the Russian Foundation for Fundamental
Research (project no. 08-02-01033-a).

%
%%%%%%%%%%%%%%%%%%%%%%%%%%%%%%%%%%%%%%%%%%%%%%%%%%%%%%%%%%%%%%%%%%%%%%%%
%
\section*{References}

\hangindent=5mm
\noindent
1 Severny, A.B.,
   {\em Ann. Rev. As\-tron. Astro\-phys.\/}, {2} (1964), 363.

\hangindent=5mm
\noindent
2 Lin, Y., Wei, X., Zhang, H.,
   {\em Solar Phys\/}., {148} (1993), 133.

\hangindent=5mm
\noindent
3 Wang, J., ,
   {\em Fun\-da\-men. Cosmic Phys\/}., {20} (1999), 251.

\hangindent=5mm
\noindent
4 Liu, C., Deng, N., Liu, Y.,
  Falconer, D., Goode, P.R., Denker, C., Wang, H.,
   {\em Astro\-phys. J\/}., {622} (2005), 722.

\hangindent=5mm
\noindent
5 Wang, H., Liu, C., Deng, Y., Zhang, H.,
   {\em Astro\-phys. J\/}., {627} (2005), 1031.

\hangindent=5mm
\noindent
6 Sudol, J.J., Harvey, J.W.,
   {\em Astro\-phys. J\/}., {635} (2005), 647.

\hangindent=5mm
\noindent
7 Gorbachev, V.S., Somov, B.V.,
   {\em Soviet Astronomy--AJ\/}, {33} (1989), 57.

\hangindent=5mm
\noindent
8 Gorbachev, V.S., Somov, B.V.,
   {\em Adv. Space Res\/}., {10} (1990), No.~9, 105.

\hangindent=5mm
\noindent
9 Gorbachev, V.S., So\-mov, B.V.,
   {\em Solar Phys\/}., {117} (1988), 77.

\hangindent=5mm
\noindent
10 Xiao, C.J., Wang, X.G., Pu, Z.Y., Ma, Z.W., Zhao, H., Zhou, G.P.,
   Wang, J.X., Kivelson, M.G., Fu, S.Y., Liu, Z.X., Zong, Q.G.,
   Dunlop, M.W., Glassmeier, K.-H., Lucek, E., Reme, H.,
   Dandouras, I., Escoubet, C.P.,
   {\em Nature Physics\/}, {3} (2007), 609.

\hangindent=5mm
\noindent
11 Dubrovin, B.A., Novikov, S.P., Fomenko, A.T.
   {\em Modern Geometry\/}, Nauka (in Russian), Moscow
   (1986).

\hangindent=5mm
\noindent
12 Syrovatskii, S.I.,
   {\em Soviet Astronomy--AJ\/}, {6} (1962), 768.

\hangindent=5mm
\noindent
13 Brushlinskii, K.V., Zaborov, A.M., Syrovatskii, S.I.,
   {\em Soviet J. Plasma Physics}, {6} (1980), 165.

\hangindent=5mm
\noindent
14 Biskamp, D.,
   {\em Phys. Fluids\/}, {29} (1986), 1520.

\hangindent=5mm
\noindent
15 Biskamp, D. {\em Nonlinear Magnetohydrodynamics\/},
    Cambridge Univ. Press, Cambridge, UK (1997).

\hangindent=5mm
\noindent
16 Yamada, M., Ren, Y., Ji, H.,
   Gerhardt, S., Dorfman, S., McGeehan, B.,
   AGU, FM, abs. SH43A-03 (2007).

\hangindent=5mm
\noindent
17 Syrovatskii, S.I.,
   {\em Ann. Rev. Astron. Astrophys.\/}, {19} (1981), 163.

\hangindent=5mm
\noindent
18
Longcope, D.W., Cowley, S.C.,
   {\em Phys. Plasmas\/}, {3} (1996), 2885.

\hangindent=5mm
\noindent
19
Somov, B.V.
   {\em Plasma Astrophysics, Part II, Reconnection and Flares\/},
   Springer, New York (2006).

\hangindent=5mm
\noindent
20
Cassak, P.A., Drake, J.F., Shay, M.A., Eckhardt, B.,
   {\em Phys. Rev. Lett\/}., {98} (2007), No.~21, id. 215001. %%% ? %%%

\hangindent=5mm
\noindent
21
Uzdensky, D.A.,
   {\em Phys. Rev. Lett\/}., {99} (2007), No.~26, id. 261101. %%% ? %%%

\hangindent=5mm
\noindent
22
Greene, J.M.,
   {\em J. Geophys. Res\/}., {93} (1988), 8583.

\hangindent=5mm
\noindent
23
Lau, Y.T., Finn, J.M.,
   {\em Astrophys. J\/}., {350} (1990), 672.

\hangindent=5mm
\noindent
24
Rust, D.M., Somov, B.V.,
   {\em Solar Phys\/}., {93} (1984), 95.

\hangindent=5mm
\noindent
25
Sweet, P.A.,
   {\em Ann. Rev. Astron. Astrophys.\/}, {7} (1969), 149.

\hangindent=5mm
\noindent
26
Somov, B.V.
   {\em Physical Processes in Solar Flares\/},
   Kluwer Academic Publ., Dordrecht (1992).

\hangindent=5mm
\noindent
27
Sakai, J.I., de Jager, C.,
   {\em Space Sci. Rev\/}., {77} (1996), 1.

\hangindent=5mm
\noindent
28
Gold, T., Hoyle, F.,
   {\em Monthly Not. Royal Astron. Soc\/}., {120} (1960), 89.

\hangindent=5mm
\noindent
29
Pevtsov, A.A., Canfield, R.C., Zi\-rin, H.,
   {\em Astro\-phys. J\/}., {473} (1996), 533.

\hangindent=5mm
\noindent
30
Den, O.G., Somov, B.V.,
   {\em Soviet Astronomy--AJ\/}, {33} (1989), 149.

\hangindent=5mm
\noindent
31
Antiochos, S.K.,
   {\em Astrophys. J\/}., {502} (1998), L181.

\hangindent=5mm
\noindent
32
Parnell, C.E.,
   {\em Solar Phys\/}., {242} (2007), 21.

\hangindent=5mm
\noindent
33
Gorbachev, V.S., Kel'ner, S.R., Somov, B.V., Shwarz, A.S.,
   {\em Soviet Astronomy--AJ\/}, {32} (1988), 308.

\hangindent=5mm
\noindent
34
Barnes, G.,
   {\it Astrophys. J\/}., {670} (2007), L53.

\hangindent=5mm
\noindent
35
Ugarte-Urra, I., Warren, H.P., Winebarger, A.R.,
   {\em Astrophys. J\/}., {662} (2007), 1293.

\hangindent=5mm
\noindent
36
Syrovatskii, S.I., Somov, B.V.
   in
   M. Dryer and E. Tandberg-Hanssen (eds.),
   {\em Solar and Interplanetary Dynamics\/},
   IAU Symp. {91},
   Reidel, Dordrecht (1980),
   pp.~425-441.

\hangindent=5mm
\noindent
37
Syrovatskii, S.I.,
   {\em Solar Phys\/}., {76} (1982), 3.

\hangindent=5mm
\noindent
38
Landau, L.D., Lifshitz, E.M., Pitaevskii, L.P.
    {\em Electrodynamics of Continuous Media\/},
    Pergamon Press, Oxford (1984).

\hangindent=5mm
\noindent
39
Kadomtsev, B.B.
   {\em Collective Phenomena in Plasma\/},
   Nau\-ka (in Russian), Moscow (1976).

\hangindent=5mm
\noindent
40
Artsimovich, L.A., Sagdeev, R.Z.
   {\em Plasma Physics for Physicists\/},
   Atomizdat, Moscow (1979).

\hangindent=5mm
\noindent
41
Syrovatskii, S.I.
   in
   E.R. Dyer (ed.),
   {\em Solar-Terrestrial Physics 1970, Part~1\/},
   D. Reidel Publ., Dordrecht, Holland (1972), pp.~119-133.

\hangindent=5mm
\noindent
42
Somov, B.V., Kosugi, T.,
   {\em Astrophys. J}., {485} (1997), 859.

\hangindent=5mm
\noindent
43
Somov, B.V., Bogachev, S.A.,
   {\em Astronomy Lett\/}., {29} (2003), 621.

\hangindent=5mm
\noindent
44
Giuliani, P., Neukirch, T., Wood, P.,
   {\em Astrophys. J\/}., {635} (2005), 636.

\hangindent=5mm
\noindent
45
Benz, A.
   {\em Plasma Astrophysics:
   Kinetic Processes in Solar and Stellar Coronae\/},
   Kluwer Academic Publ., Dordrecht (2002).

\hangindent=5mm
\noindent
46
Masuda, S., Kosugi, T., Hudson, H.S.,
   {\em Solar Phys\/}., {204} (2001), 57.

\hangindent=5mm
\noindent
47
Lin, R.P., Krucker, S., Hur\-ford, G.J.,
   Smith, D.M., Hudson, H.S., Hol\-man, G.D., Schwartz, R.A.,
   Dennis, B.R., Share, G.H., Murphy, R.J., Ems\-lie, A.G.,
   Johns-Krull, C., Vil\-mer, N.,
   {\em Astro\-phys. J\/}., {595} (2003), L69.

\hangindent=5mm
\noindent
48
Kosugi, T., Dennis, B.R., Kai, K.,
   {\em Astrophys. J\/}., {324} (1988), 1118.

\hangindent=5mm
\noindent
49
Qiu, J., Lee, J., Gary, D.E.,
   {\em Astrophys. J\/}., {603} (2004), 335.

\hangindent=5mm
\noindent
50
Li, Y.P., Gan, W.Q.,
   {\em Astrophys. J\/}., {629} (2005), L137.

\hangindent=5mm
\noindent
51
Somov, B.V.,
   {\em Soviet Phys. Usp\/}., {28} (1985), 271.

\hangindent=5mm
\noindent
52
Bagal\'{a}, L.G., Mandrini, C.H., Rovira, M.G.,
D\'{e}\-mou\-lin, P., H\'{e}\-noux, J.C.,
   {\em Solar Phys}., {161} (1995), 103.

\hangindent=5mm
\noindent
53
Aschwanden, M.J., Kosugi, T., Hanaoka, Y., Nishio, M., Melrose, D.,
   {\it Astrophys. J\/}., {526} (1999), 1026.

\hangindent=5mm
\noindent
54
Morita, S., Uchida, Y., Hirose, S., Uemura, S., Yamaguchi, T.,
   {\em Solar Phys\/}., {200} (2001), 137.

\hangindent=5mm
\noindent
55
Longcope, D.W., McKenzie, D.E., Cir\-tain, J., Scott, J.,
   {\em Astrophys. J\/}., {630} (2005), 596.

\hangindent=5mm
\noindent
56
Longcope, D.W., Beveridge, C.,
   {\em Astrophys. J\/}., {669} (2007), 621.

\hangindent=5mm
\noindent
57
Somov, B.V.,
   {\em Astron. Astrophys\/}., {163} (1986), 210.

\hangindent=5mm
\noindent
58
Priest, E.R. and For\-bes, T.
   {\em Magnetic Reconnection:
   MHD Theory and Applications\/},
   Cambridge Univ. Press, Cambridge, UK (2000).

\hangindent=5mm
\noindent
59
Syrovatskii, S.I.,
   {\em Soviet Physics--JETP\/}, {23} (1966), 754.

\hangindent=5mm
\noindent
60
Horiuchi, R., Pei, W., Sato, T.,
   {\em Earth Planets Space\/}, {53} (2001), 439.

\hangindent=5mm
\noindent
61
Joshi, B., Manoharan, P.K., Veronig, A.M., Pant, P., Pandey, K.,
   {\em Solar Phys\/}., {242} (2007), 143.

\hangindent=5mm
\noindent
62
Liu, W., Petrosian, V., Dennis, B.R., Jiang, Y.W.,
   {\em Astro\-phys. J\/}., {676} (2008), 704.

\hangindent=5mm
\noindent
63
Jain, R., Pradhan, A.K., Joshi, V., Shah, K.J., Trivedi, J.J.,
   Kayasth, S.L., Shah, V.M., Deshpande, M.R.,
   {\em Solar Phys\/}., {239} (2006), 217.

\hangindent=5mm
\noindent
64
Jain, N., Sharma, A.S.,
   AGU, FM, abs. SM31D-0659 (2007).

\hangindent=5mm
\noindent
65
Stark, D., Sharma, A., Jain, N.,
   AGU, FM, abs. SM31C-1465 (2007).

\hangindent=5mm
\noindent
66
Litvinenko, Y.E., Somov, B.V.,
   {\em Solar Phys\/}., {146} (1993), 127.

\hangindent=5mm
\noindent
67
Litvinenko, Y.E.; Somov, B.V.,
   {\em Solar Phys\/}., {158} (1995), 317.

\hangindent=5mm
\noindent
68
Karlick\'y, M., Kosugi, T.,
   {\em Astron. Astrophys\/}., {419} (2004), 1159.

\hangindent=5mm
\noindent
69
Bogachev, S.A., Somov, B.V.,
   {\em Astronomy Letters\/}, {33} (2007), 54.

\hangindent=5mm
\noindent
70
Somov, B.V.,
   {\em Astronomy Letters\/}, {34} (2008), 635.

%%%%%%%%%%%%%%%%%%%%%%%%%%%%%%%%%%%%%%%%%%%%%%%%%%%%%%%%%%%%%%%%%%%%%%%%

\end{document}